\documentclass[usegraphicx,usenatbib]{mn2e}
\usepackage{times}
\usepackage{array}

\title[Bright stars observed by \textit{FIMS}/\textit{SPEAR}]{Bright stars observed by
FIMS/SPEAR}

\author[Jo et al.]
  {Young-Soo Jo,$^{1,2}$\thanks{E-mail: stspeak@kasi.re.kr}
  Kwang-Il Seon,$^{1,3}$
  Kyoung-Wook Min,$^2$
  Yeon-Ju Choi,$^{2,4}$
  \newauthor 
  Tae-Ho Lim,$^2$
  Yeo-Myeong Lim,$^2$
  Jerry Edelstein,$^5$
  and Wonyong Han$^1$\\
  $^1$Korea Astronomy and Space Science Institute (KASI), 776 Daedeokdae-ro, Yuseong-gu, Daejeon, Korea 305-348, Republic of Korea\\
  $^2$Korea Advanced Institute of Science and Technology (KAIST), 291 Daehak-ro, Yuseong-gu, Daejeon, Korea 305-701, Republic of Korea\\
  $^3$Astronomy and Space Science Major, Korea University of Science and Technology, 217 Gajeong-ro, Yuseong-gu, Daejeon, Korea 305-350, Republic of Korea\\
  $^4$Korea Aerospace Research Institute (KARI), 169-84 Gwahak-ro, Yuseong-gu, Daejeon, Korea 305-806, Republic of Korea\\
  $^5$Space Sciences Laboratory, University of California, Berkeley, CA, USA}

\begin{document}

\pagerange{\pageref{firstpage}--\pageref{lastpage}} \pubyear{2015}

\maketitle

\label{firstpage}

\begin{abstract}
In this paper, we present a catalogue of the spectra of bright stars
observed during the sky survey using the Far-Ultraviolet Imaging
Spectrograph (\textit{FIMS}), which was designed primarily to
observe diffuse emissions. By carefully eliminating the
contamination from the diffuse background, we obtain the spectra of
70  bright stars observed for the first time with a spectral
resolution of 2--3 {\AA} over the wavelength of 1370--1710 {\AA}.
The far-ultraviolet spectra of an additional 139  stars are also
extracted with a better spectral resolution and/or higher
reliability than those of the previous observations. The stellar
spectral type of the stars presented in the catalogue spans from O9
 to A3. The method of spectral extraction of the bright
stars is validated by comparing the spectra of 323 stars with those
of the International Ultraviolet Explorer (\textit{IUE})
observations.
\end{abstract}


\section{Introduction}

Among the space missions that have observed stellar spectra in the
range of the far ultraviolet (FUV) wavelength, the International
Ultraviolet Explorer \citep[\textit{IUE};][]{bog78} has the most
extensive list of targets with approximately 5,000 stars with their
spectra taken for the wavelength range of 1150--1980 {\AA} with a
resolution of 0.1--0.3 {\AA}. Although the wavelength coverage might
differ slightly among various missions, there have been a number of
additional small and full-scale missions. These include the S2/68
Ultraviolet Sky Survey Telescope
\citep[\textit{UVSST};][]{bok73,jam76} aboard the ESRO Satellite
TD-1, Copernicus \citep{rog73,sno77}, \textit{SKYLAB} Experiment
S-019 \citep{hen75,hen79}, Hopkins Ultraviolet Telescope
\citep[\textit{HUT};][]{dav92,kru95}, Orbiting Retrievable Far and
Extreme Ultraviolet Spectrometers \citep[\textit{ORFEUS};][]{jen96},
Far Ultraviolet Spectroscopic Explorer
\citep[\textit{FUSE};][]{moo00,sah00}, Galaxy Evolution Explorer
\citep[\textit{GALEX};][]{mor07,ber11}, and  Space Telescope Imaging
Spectrograph (\textit{STIS}; \citealt{woo98} and \textit{StarCAT};
\citealt{ayr10}), Goddard High Resolution Spectrograph
\citep[\textit{GHRS};][]{bra94}, Cosmic Origins Spectrograph
\citep[\textit{COS};][]{gre12}  aboard the Hubble Space Telescope
(HST). As a result of these observations, the spectra of
approximately 10,000 stars are now available in the FUV wavelengths.

Recently, the Far-Ultraviolet Imaging Spectrograph (\textit{FIMS}),
also known as the Spectroscopy of Plasma Evolution from
Astrophysical Radiation (\textit{SPEAR}), performed an all-sky
survey in the FUV wavelength region. The \textit{FIMS} is a dual
channel imaging spectrograph (\textit{S}-band: 900--1150 {\AA},
\textit{L}-band: 1370--1710 {\AA}) with moderate spectral
($\lambda/\Delta\lambda$ $\sim$ 550) and angular ($\sim$ 5\arcmin)
resolutions, and it was designed for the observation of diffuse
emissions from interstellar medium \citep{ede06a,ede06b}. The
primary purpose of the \textit{FIMS} is to study the FUV emission
from atomic, ionic, and molecular species in a variety of
interstellar environments \citep[e.g.][]{kor06}. In addition, a
number of bright stars have also been observed when they are within
its large image fields of view, defined by the slit size
(\textit{S}-band: 4\fdg0 $\times$ 4\farcm6, \textit{L}-band: 7\fdg5
$\times$ 4\farcm3) . This paper discusses the spectral extraction of
these observed bright stars.

The \textit{FIMS} was used to make observations for one and a half
years, covering more than 80\% of the sky, after its launch by the
Korean microsatellite \textit{STSAT-1} on 27 September 2003 into a
700 km sun-synchronous orbit. Although stars were observed in both
the \textit{S}- and \textit{L}-bands, the \textit{S}-band data were
excluded from the present analysis because of their strong
contamination with geocoronal emission lines as well as the low
detector sensitivity. We extracted the \textit{L}-band spectra for
532 stars from 1,523 orbits of observations during its mission
lifetime, from which 70  stars were observed for the first time.
Here, we report the FUV spectra of these stars along with those of
the 139  stars that were observed with a better spectral resolution
and/or higher reliability than those of the previous observations,
as the \textit{FIMS} catalogue stars. The remaining 323
 stars, which were also observed by the \textit{IUE} in
a large aperture mode , were used to validate the \textit{FIMS}
spectra. Most of the 323 stars were observed in the high-dispersion
mode of the \textit{IUE} with a spectral resolution of $\sim$0.2
{\AA}, but 34 stars among them were observed in the low-dispersion
mode and with a lower spectral resolution of $\sim$6 {\AA}.
 Section 2 describes the data processing steps used to
obtain the FUV spectra of the observed stars, and Section 3 presents
the detailed descriptions of the statistical properties of the
\textit{FIMS} catalogue stars. A summary is provided in Section 4.

\section{Data Processing Steps}

We followed three key steps to obtain the spectral information of
the stars that were observed with a diffuse background. A detailed
description about flat-fielding as well as the procedures of
wavelength and flux calibration for the \textit{FIMS} data can be
found in \citet{ede06b}. In the present paper, we report only on the
process of spectral extraction of bright stars from the existing
\textit{FIMS} archival data. However, the new effective area was
derived as it was seen to change significantly over the mission time
due to degradation in detector sensitivity (Section 2.3).

\begin{figure}
 \begin{center}
    \includegraphics[width=3.8cm]{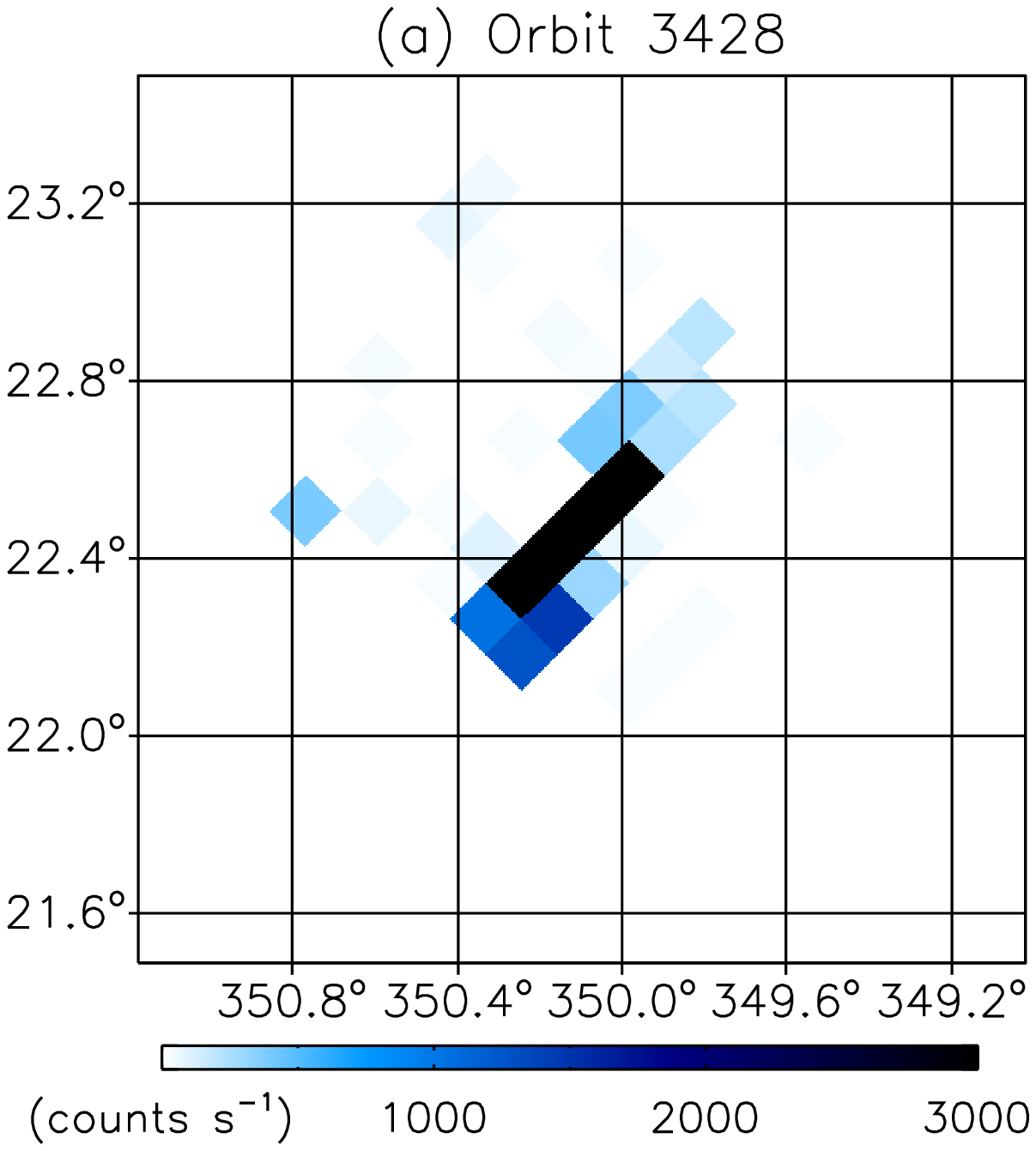}\quad\quad
    \includegraphics[width=3.8cm]{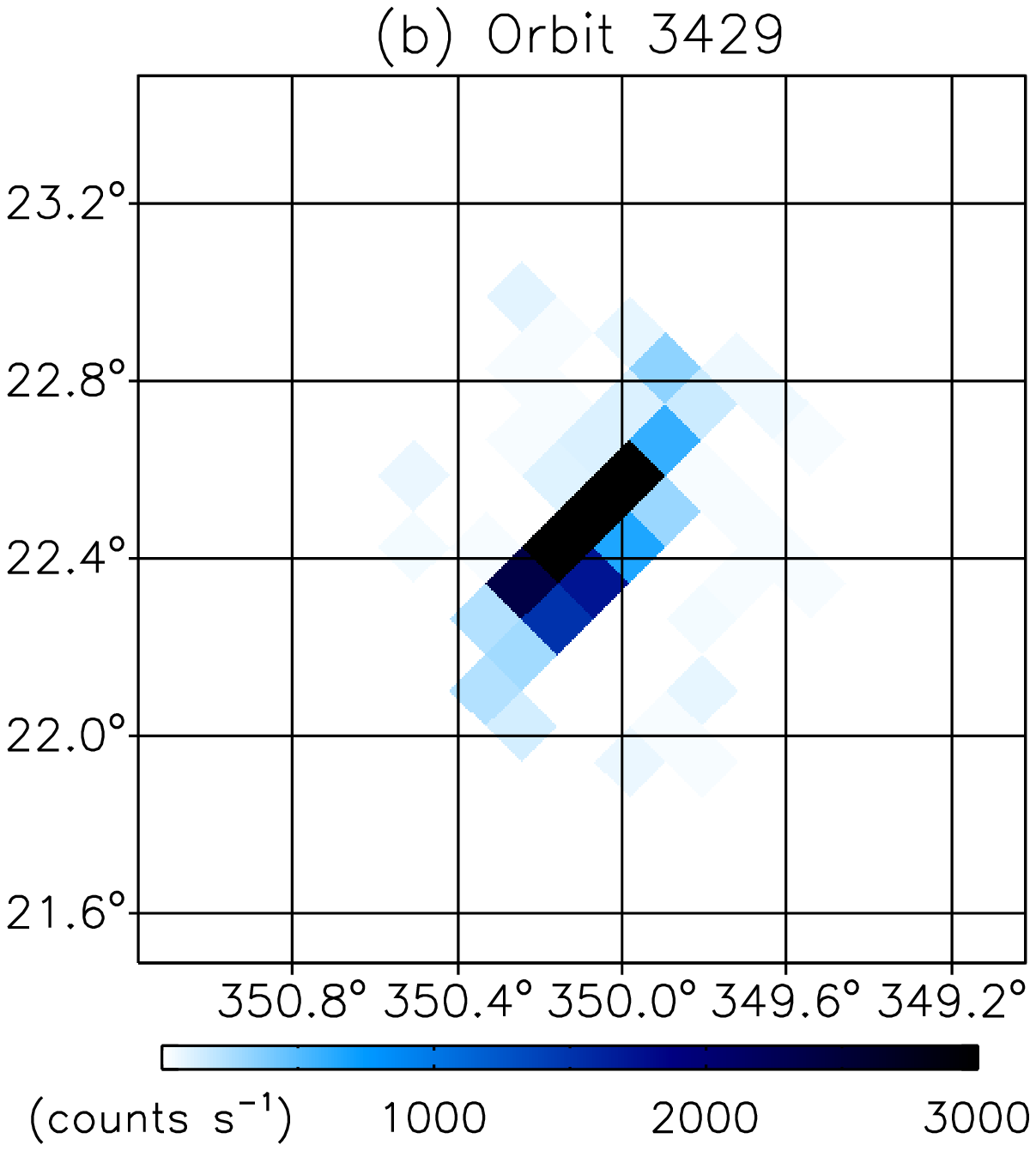}\\
    \vspace{10pt}
    \includegraphics[width=3.8cm]{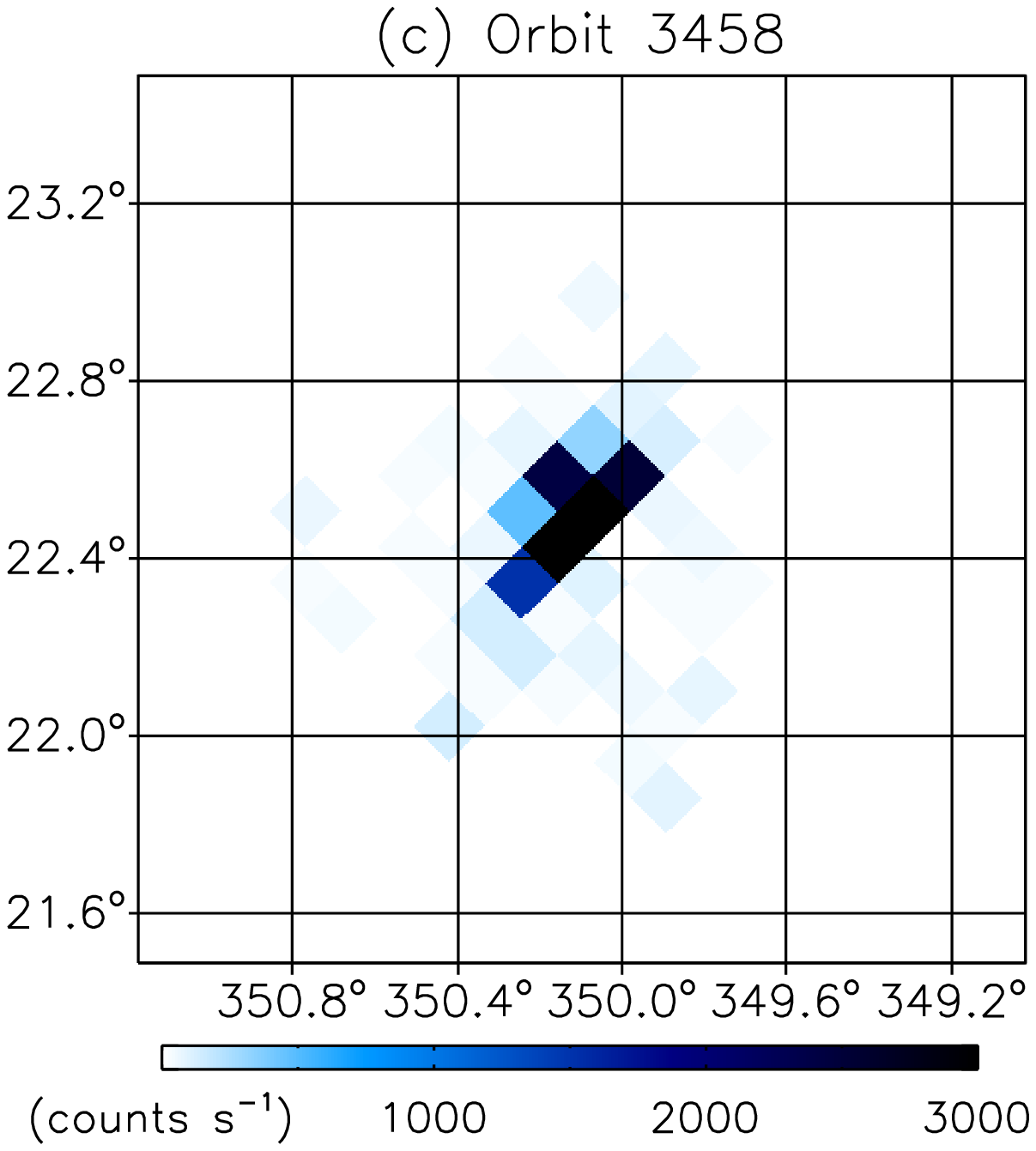}\quad\quad
    \includegraphics[width=3.8cm]{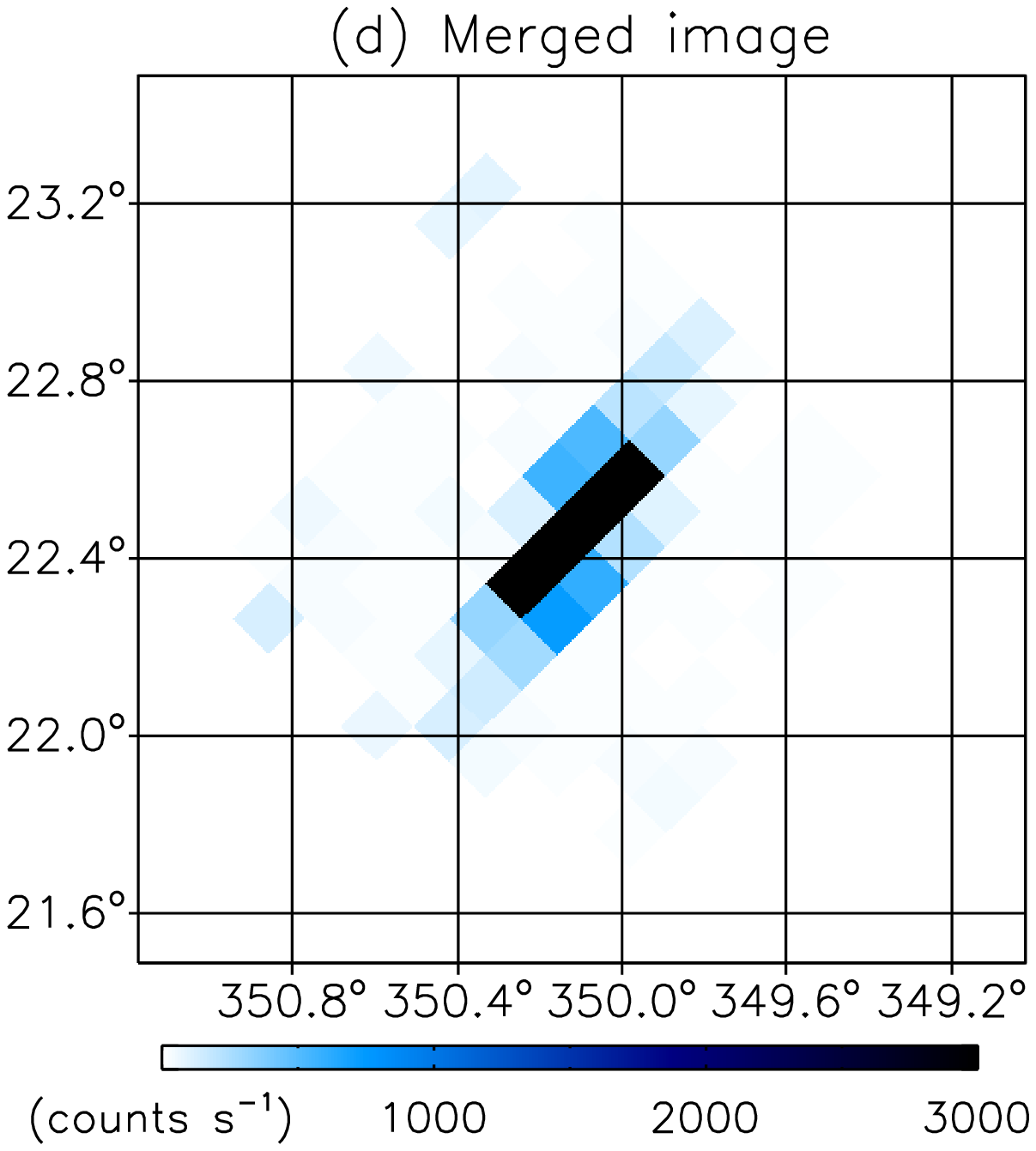}
 \end{center}
 \caption{Photon count-rate (unit: count s$^{-1}$) images of HD 143275 with Galactic coordinates:
 the multiple images of (a) to (c)  are merged into one,
 as shown in (d) . \label{fig:tile}}
\end{figure}

\begin{figure}
 \begin{center}
  \includegraphics[width=3.8cm]{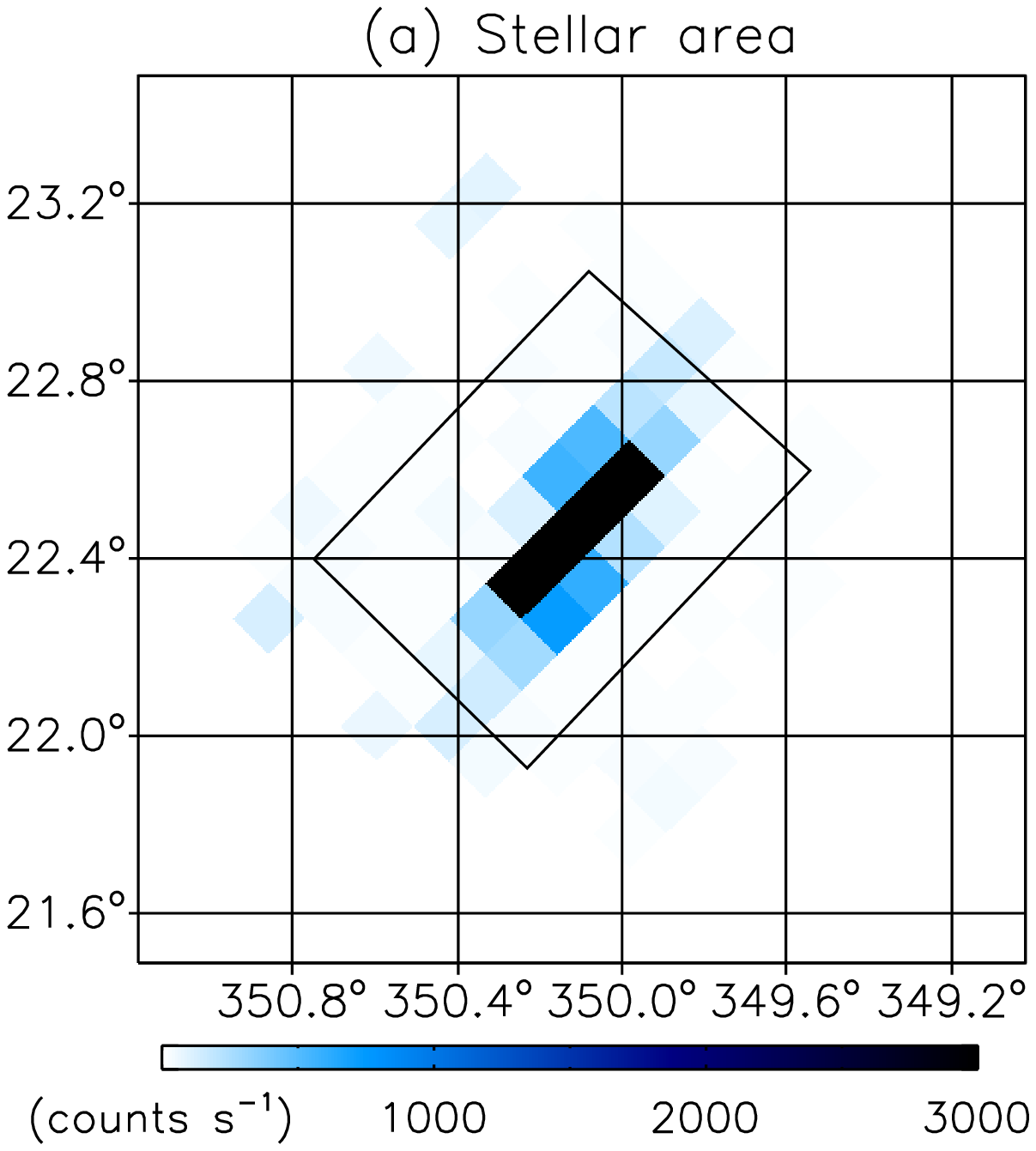}\quad\quad
  \includegraphics[width=3.8cm]{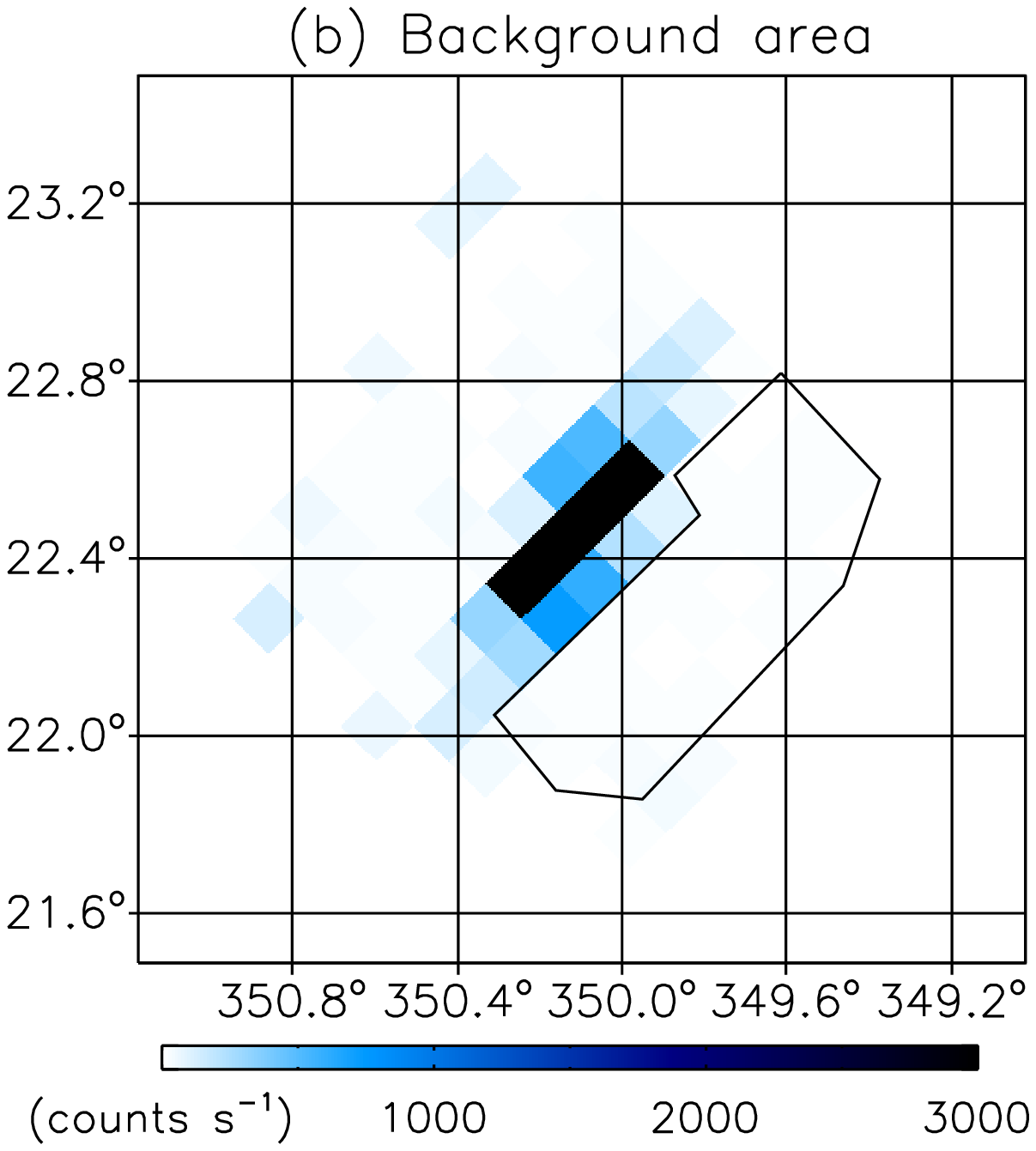}\\
 \end{center}
 \caption{Stellar and background areas selected for HD 143275 are
 presented in the left and right panels, respectively. \label{fig:tile_area}}
\end{figure}

\subsection{Step 1: Identification of stars and merging of data}

The \textit{FIMS} data, which were archived in the FITS format with
coordinates and wavelengths assigned to each photon, were arranged
to form an image tile for each orbit of observations and pixelated
using the {\sc HEALPix} scheme \citep{gor05} with a resolution
parameter of \textit{Nside} = 512, corresponding to a pixel size of
approximately 7 arcmin. The constant pixel area of the {\sc HEALPix}
scheme makes it convenient to extract stellar spectra, which are
mixed with the background spectra because of instrumental
scattering.

We identified the stars by comparing the \textit{FIMS} images with
the locations of the bright stars in the \textit{TD1} star catalogue
\citep{tho78} based on the S2/68 Ultraviolet Sky Survey Telescope
(\textit{UVSST}) of the \textit{TD1} satellite. \textit{UVSST} is
described in \citet{bok73} and the absolute calibration of the
instrument is given in \citet{hum76}. \textit{UVSST} consisted of an
f/3.5 telescope with a primary mirror having a diameter of 27.5 cm,
feeding photons to a spectrometer and photometer. The spectrometer
has an entrance aperture of 11\farcm8 $\times$ 17\arcmin and a
wavelength band in the range of 130--255 nm with spectral resolution
of 35 {\AA}. The photometer has an aperture of 1\farcm7 $\times$
17\arcmin with a broad passband (31 nm) centred at 274 nm. The first
\textit{UVSST} spectral catalogue for 1,356 stars was published by
\citet{jam76}\footnote{http://vizier.u-strasbg.fr/viz-bin/VizieR?-source=III/39A}.
Later,
\citet{tho78}\footnote{http://vizier.u-strasbg.fr/viz-bin/VizieR?-source=II/59}
extended the catalogue to 31,215 stars and provided the absolute UV
fluxes in four passbands: 135--175 nm, 175--215 nm, 215--255 nm, and
the photometric band at 274 nm. The fluxes of the first three
wavelength bands were obtained by binning the spectrophotometric
data, and the flux of the longest wavelength band was based on the
photometric data only. Hence, the angular resolution of the
catalogue is limited by that of the photometer, which is 2 arcmin.
It is notable that the shortest passband of 135--175 nm, centred at
1565 {\AA} (henceforth, F1565), is comparable with that of the
\textit{FIMS} \textit{L}-band and was used for identification of UV
bright stars in the present study because the \textit{TD1} is more
sensitive and has higher angular resolution than \textit{FIMS}. We
note that the photometric catalogue (henceforth, \textit{TD1})
extended by \citet{tho78} were used to identify stars, while the
spectral catalogue (henceforth, \textit{UVSST}) made by
\citet{jam76} were used to compare with the \textit{FIMS} spectra
after Section 3.

In order to avoid misidentification of stars in crowded areas and
contamination by other bright stars in the surrounding region, we
selected only the stars listed in the \textit{TD1} catalogue that
were isolated within the 2\degr $\times$ 2\degr\ angular region
centred around the target stars. Since the angular resolution of
\textit{TD1} (2\arcmin) is higher than that of the \textit{FIMS}, we
believe that the \textit{TD1} catalogue is able to resolve the stars
even for those that the \textit{FIMS} cannot if the stars were
bright enough to be observed by \textit{TD1}. Hence, the angular
resolution of the \textit{FIMS} catalogue is limited by that of the
\textit{TD1} catalogue. Further, we checked the 2\degr $\times$
2\degr\ \textit{FIMS} images and discarded the pixels brighter than
the background median values by a factor of three, except those
associated with the target stars. The faintest star in the resulting
\textit{FIMS} catalogue has a flux of $\sim$7 $\times$ 10$^{-12}$
erg s$^{-1}$ cm$^{-2}$ {\AA}$^{-1}$, much higher than the typical
\textit{TD1} flux limit of 10$^{-12}$ erg s$^{-1}$ cm$^{-2}$
{\AA}$^{-1}$ in the spectral band of 1350--1750 {\AA}. Although we
avoided regions of high concentration of UV sources on the basis of
the \textit{TD1}, there may still exist some stars which were not
resolved in the \textit{TD1} catalogue among the \textit{FIMS}
catalogue stars. However, we found no bright stars near the stars
included in the \textit{FIMS} catalogue. Instead, some stars that
were observed by \textit{FIMS} but not included in our catalogue
were found to be contaminated by unresolved stars. For example, HD
214167 \citep[B1.5Vs;][]{abt11} was found to form a double system
with one of the \textit{FIMS} star of HD 214168
\citep[B1Vne:;][]{abt11}, separated by an angular distance of
$\sim$0.37\arcmin. As another example, HD 199739 with a spectral
type of B2II \citep{hil77} was found at an angular distance of
4\arcmin\ from one of the \textit{FIMS} star of HD 199661. However,
the star is fainter than HD 199661 by 1.7 mag (21\%) in V band and
2.8 mag (8\%) in U band \citep{ree03}. These two stars were not
included in the \textit{FIMS} catalogue since they were also
observed by the IUE. In summary, although the 2\degr $\times$
2\degr\ angular region may contain many other stars not resolved in
the \textit{TD1}, the contamination due to them would not be
significant.

\begin{figure*}
 \begin{center}
  \includegraphics[width=15cm]{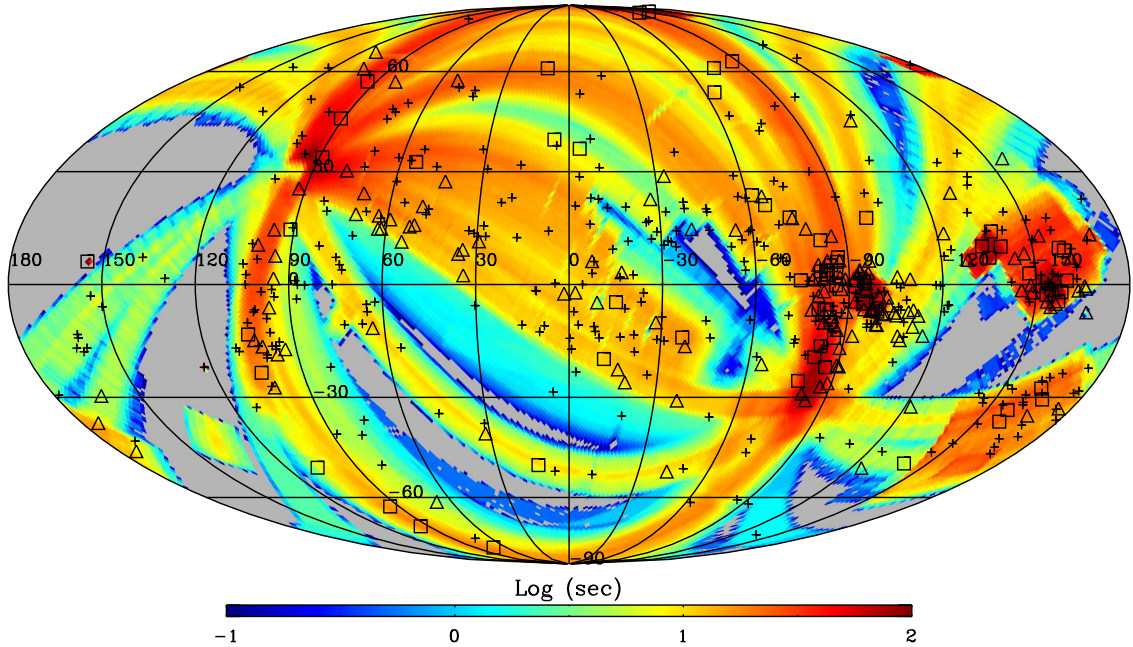}
 \end{center}
 \caption{Galactic locations of the 532 stars observed by \textit{FIMS} plotted
 on an all-sky exposure time map. The squares indicate the 70  stars
 observed for the first time by \textit{FIMS}, and the crosses indicate the 323
 stars observed by \textit{IUE}. The remaining 139  stars are represented by
 triangles; these were observed in other missions such as \textit{UVSST} and \textit{SKYLAB}.
 \label{fig:allsky}}
\end{figure*}

Once the target star was identified and the region of 2\degr
$\times$ 2\degr\ tiles were confirmed to be free of bright stars, we
performed a Gaussian fitting for each image to find the centre of
the star. We note that instrumental scattering results in diffuse
images even for a point source. The selection of this large area
around each star is necessary because the instrumental scattering by
the slit caused the elongated image of a bright star along the slit
direction, as shown in Figure \ref{fig:tile} for HD 143275. The
multiple image tiles produced by multiple observations for the same
star were merged into one before spectral information was extracted.
The range of number of tiles for each star was from 3 to 207, with
an average of 45.

\begin{table}
 \centering
  \caption{List of the \textit{IUE} reference stars, together
  with the orbit number and the dates of the \textit{FIMS} observations. \label{tbl:eff}}
  \begin{tabular}{ccc|ccc}
  \hline
  HD ID & Orbit & Observational & HD ID & Orbit & Observational\\
        &       & date          &       &       & date\\
  \hline
  21790 & 1116 & 2003-12-12 & 100889 & 2559 & 2004-03-20 \\
  46487 & 1254 & 2003-12-21 & 83058  & 2561 & 2004-03-20 \\
  32249 & 1292 & 2003-12-24 & 79447  & 2772 & 2004-04-03 \\
  31726 & 1341 & 2003-12-27 & 98718  & 2973 & 2004-04-17 \\
  25340 & 1455 & 2004-01-04 & 105937 & 3053 & 2004-04-22 \\
  68217 & 1929 & 2004-02-05 & 108257 & 3054 & 2004-04-22 \\
  63922 & 1964 & 2004-02-08 & 152614 & 3610 & 2004-05-30 \\
  64740 & 1994 & 2004-02-10 & 158094 & 3847 & 2004-06-16 \\
  70556 & 2006 & 2004-02-11 & 165024 & 3947 & 2004-06-22 \\
  64802 & 2057 & 2004-02-14 & 166182 & 3955 & 2004-06-23 \\
  76566 & 2070 & 2004-02-15 & 172910 & 4061 & 2004-06-30 \\
  75821 & 2075 & 2004-02-15 & 207971 & 4639 & 2004-08-09 \\
  67797 & 2078 & 2004-02-16 & 188665 & 5226 & 2004-09-18 \\
  76805 & 2443 & 2004-03-12 & 207330 & 5635 & 2004-10-16 \\
  \hline
\end{tabular}
\end{table}

\begin{figure*}
 \begin{center}
  \includegraphics[width=15cm]{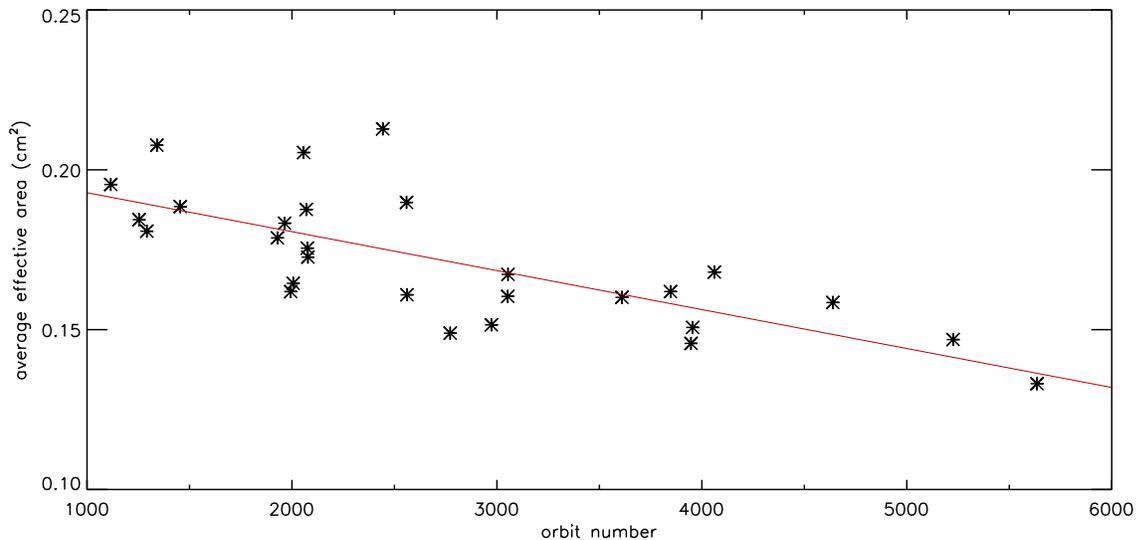}
 \end{center}
 \caption{Effective areas calculated for the \textit{IUE} reference stars
 listed in Table \ref{tbl:eff}: they are arranged according to the orbit numbers in which
 the stars were observed by the \textit{FIMS}. The solid red line is a linear fit of the data.
 \label{fig:eff_fix}}
\end{figure*}

\begin{figure*}
 \begin{center}
  \includegraphics[width=15cm]{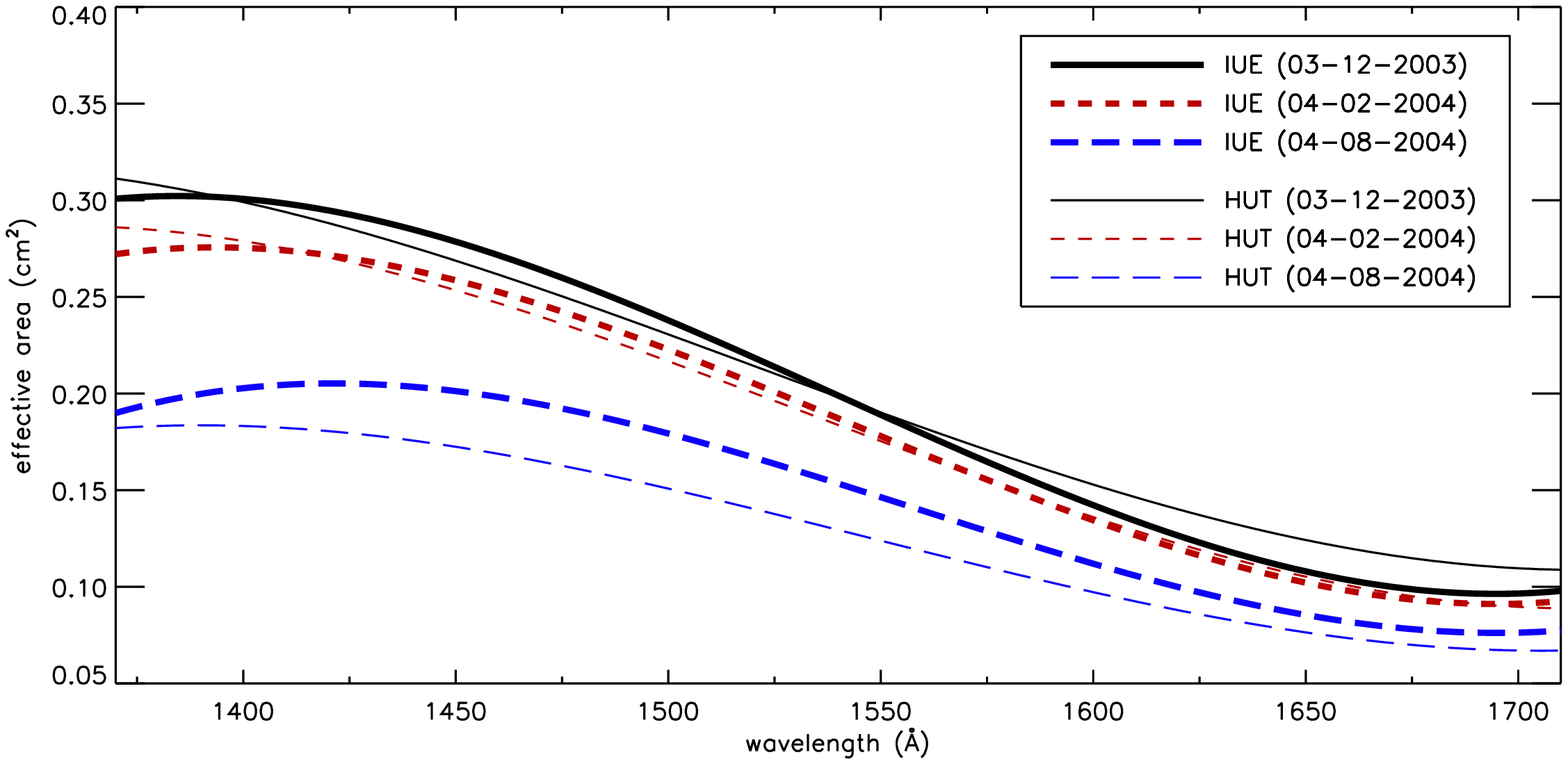}
 \end{center}
 \caption{The \textit{IUE}-based effective areas are compared with
 those of the previous \textit{HUT}-based effective areas for three representative dates.
 The thick solid black, dashed red, and the long-dashed blue lines represent the \textit{IUE}-based
 calibrations and the corresponding thin lines are the \textit{HUT}-based calibrations,
 for the dates of 03-12-2003, 04-02-2004, and 04-08-2004, respectively.
 \label{fig:eff_new}}
\end{figure*}

\subsection{Step 2: Extraction of the stellar spectrum}

The final merged 2\degr $\times$ 2\degr image tiles consist of two
areas: one for the star (hereafter, the stellar area), where both
the stellar and diffuse background photons coexist, and the other
for the background (hereafter, the background area), where the
photons from only the diffuse background are contained. We note that
the diffuse background may also contain flux from unresolved fainter
stars, although we have carefully filtered out bright background
pixels through the Step 1. The selection of these two areas was
performed manually so that the background area was free of stellar
photons of the target star.
 Figure \ref{fig:tile_area} presents an example of the
stellar area and background area selected for HD 143275.

Because the stellar area contains photons both from the star and the
diffuse background, the photon counts per angstrom for the stellar
area are expressed as follows:
\begin{equation}
   F_S(\lambda) = S(\lambda) \cdot T_S + B(\lambda) \cdot T_S \cdot \Omega_S,
\end{equation}
 where
 $F_S(\lambda)$ is the observed photon counts per angstrom for the
 stellar area (unit: photons {\AA}$^{-1}$),
 $S(\lambda)$ is the count rate (unit: photons s$^{-1}$ {\AA}$^{-1}$)
 of the stellar photons to be extracted,
 $T_S$ is the observation time for the stellar area (unit: s),
 $B(\lambda)$ is the count rate per steradian (unit: photons s$^{-1}$
 {\AA}$^{-1}$ sr$^{-1}$) of the diffuse background photons to be
 removed from $F_S(\lambda)$, and
 $\Omega_S$ is the celestial surface area of the stellar area (unit: steradians).
 Likewise, the photon counts per angstrom for the background
 area are expressed as follows:
\begin{equation}
   F_B(\lambda) = B(\lambda) \cdot T_B \cdot \Omega_B,
\end{equation}
 where
 $F_B(\lambda)$ is the observed photon count per angstrom for the
 background area (unit: photons {\AA}$^{-1}$),
 $T_B$ is the observation time for the background area (unit: s), and
 $\Omega_B$ is the celestial surface area of the background area in units
 of steradians. Hence, the stellar photon count rate of $S(\lambda)$ can be written as:
\begin{equation}
   S(\lambda) = \frac{F_S(\lambda)}{T_S} - \frac{F_B(\lambda)}{T_B} \cdot \frac{\Omega_S}{\Omega_B},
\end{equation}

We verified the signal-to-noise ratio (SNR) of $S(\lambda)$ for each
star and selected only those with a wavelength-averaged SNR greater
than 3.0. The total number of stars selected for spectral analysis
was 532, out of approximately 4,000 stars from the \textit{FIMS}
observations that were identified through comparison with the
\textit{TD1} catalogue. Among the 532 stars, 70  stars were observed
for the first time by the \textit{FIMS}, 323  stars were observed by
\textit{IUE} with better SNRs , and the remaining 139 stars were
observed by other missions, such as \textit{UVSST} and
\textit{SKYLAB}, but with lower quality than the \textit{FIMS}.
Figure \ref{fig:allsky} indicates the locations of the 532 stars
observed by \textit{FIMS} plotted on the all-sky exposure time map.

\subsection{Step 3: Correction of effective areas}

Accurate calibration of the effective areas as a function of
wavelength is necessary in order to obtain reliable stellar spectra
because the detector sensitivity degrades over time. The
\textit{FIMS} observed standard stars three times for this purpose:
a white dwarf G191-B2B and a B5V-type star HD 188665 were observed
in December 2003, February 2004, and August 2004. These observations
were compared with the reference spectra obtained previously by
\textit{HUT}, and the effective areas were obtained \citep{kim04}.
 In order to confirm these calibration results, we used
the \textit{IUE} stars and recalculated the effective areas. The
\textit{IUE} reference stars were selected according to the
following criteria:

\begin{enumerate}
  \item Single stars,
  \item Average SNR of the \textit{FIMS} observation was greater than
  7.0,
  \item B-type stars, in which spectrum was nearly flat without
strong absorption features over the FUV wavelength range
corresponding to the \textit{FIMS} \textit{L}-band, and
  \item Flux difference between the \textit{IUE} and \textit{TD1} observations was below 5$\%$.
\end{enumerate}

O-type stars were excluded because of a large number of strong
absorption lines in the FUV passband, and A-type stars were also
excluded because of their weak fluxes at short wavelengths. B-type
stars were therefore adopted for the reference stars, though some
B-type stars (especially supergiants) may still show stellar
absorption lines. The New Spectral Image Processing System
\citep[NEWSIPS;][]{nic96} dataset of the \textit{IUE} Final Archive,
which was retrieved through the FTP
site\footnote{ftp://archive.stsci.edu/pub/iue/data}, was used in the
present analysis. In particular, the data observed with a short
wavelength camera (1150--1980 {\AA}), taken in the observational
mode with a high dispersion and large aperture, was used. A total of
28 stars were selected as the \textit{IUE} reference stars and are
listed in Table \ref{tbl:eff}.

\begin{figure}
 \begin{center}
  \includegraphics[width=8cm]{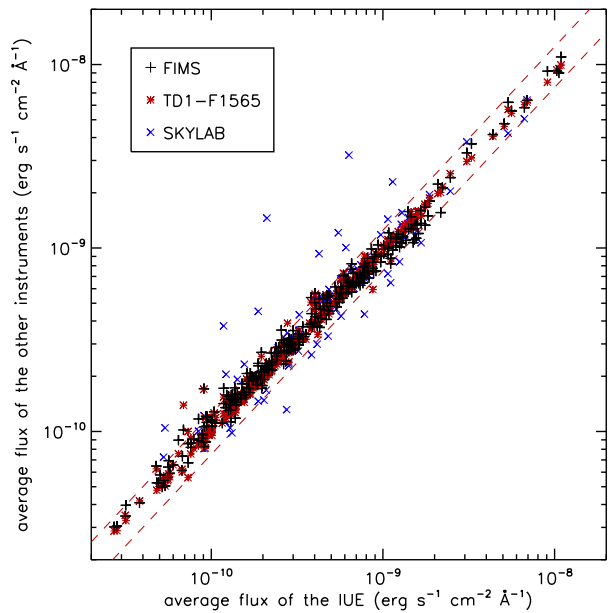}
 \end{center}
 \caption{Fluxes of the \textit{FIMS}, \textit{TD1}-F1565, and \textit{SKYLAB}
 compared with those of the \textit{IUE}. The black plus (+), red asterisk, and blue cross ($\times$)
 symbols indicate the stars observed by the \textit{FIMS}, \textit{TD1}, and \textit{SKYLAB}, respectively.
 The two dashed red lines indicate $\pm$25$\%$ boundaries of the \textit{IUE} flux.
 Both axes are log-scaled. \label{fig:flux}}
\end{figure}

We calculated the \textit{FIMS} effective areas for the 28
\textit{IUE} stars by dividing the average \textit{FIMS} photon
count rates per unit wavelength (photons s$^{-1}$ {\AA}$^{-1}$) by
the average \textit{IUE} fluxes (photons s$^{-1}$ cm$^{-2}$
{\AA}$^{-1}$) for the spectral range from 1370 to 1710 {\AA} and
plotted them as a function of the orbit number, which can be
converted into time. Figure \ref{fig:eff_fix} depicts the result,
which clearly shows that the \textit{FIMS} effective area decreased
in time. We fitted the plot using a linear function of the orbit
number, as denoted by the solid red line. Figure \ref{fig:eff_new}
presents the final results of the effective area model for three
representative orbits, which are compared with the
\textit{HUT}-based calibration. As can be seen in the figure, the
\textit{IUE}-based effective areas are in good agreement with the
\textit{HUT}-based calibration, except the one corresponding to the
4 August 2004 observation, which shows a larger effective area
compared to that of the \textit{HUT}-based calibration. However, we
believe that the \textit{IUE}-based calibration is statistically
more reliable since it is based on many stars, and the
\textit{HUT}-based calibration is based on only two stars with
limited observations \citep{kim04}. Figure \ref{fig:eff_fix} shows
fluctuations in the estimated effective area based on the stars
observed even at similar epochs, and we believe such fluctuations
are the reason for the \textit{HUT}-based curve of 4 August 2004
being lower than that based on the \textit{IUE} stars. The
fluctuations could be due to statistical variation, stellar
variability, higher Earth-background, or any instrumental failure.
We further confirmed the result of estimation of effective areas by
comparing the calibrated \textit{FIMS} fluxes averaged over
1370--1710 {\AA} with the averaged \textit{IUE} fluxes over the same
wavelength band for all of the available 323 stars, as shown in
Figure \ref{fig:flux}. It is seen that the two fluxes are in good
agreement with each other within a 25$\%$ error for more than 92$\%$
of the stars.

\begin{figure*}
 \begin{center}
  \includegraphics[width=13.4cm]{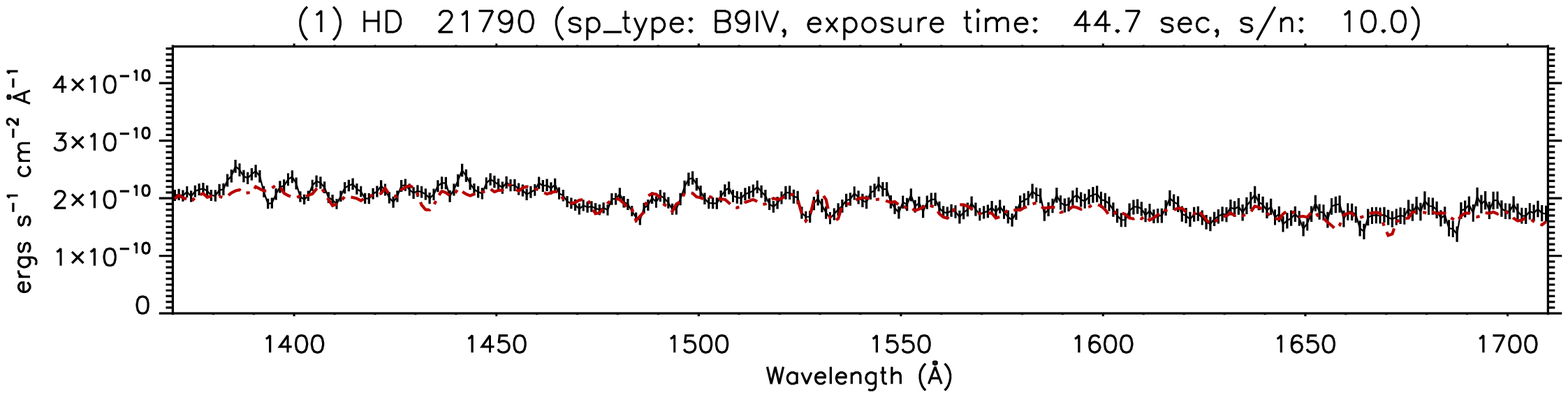}\\
  \includegraphics[width=13.4cm]{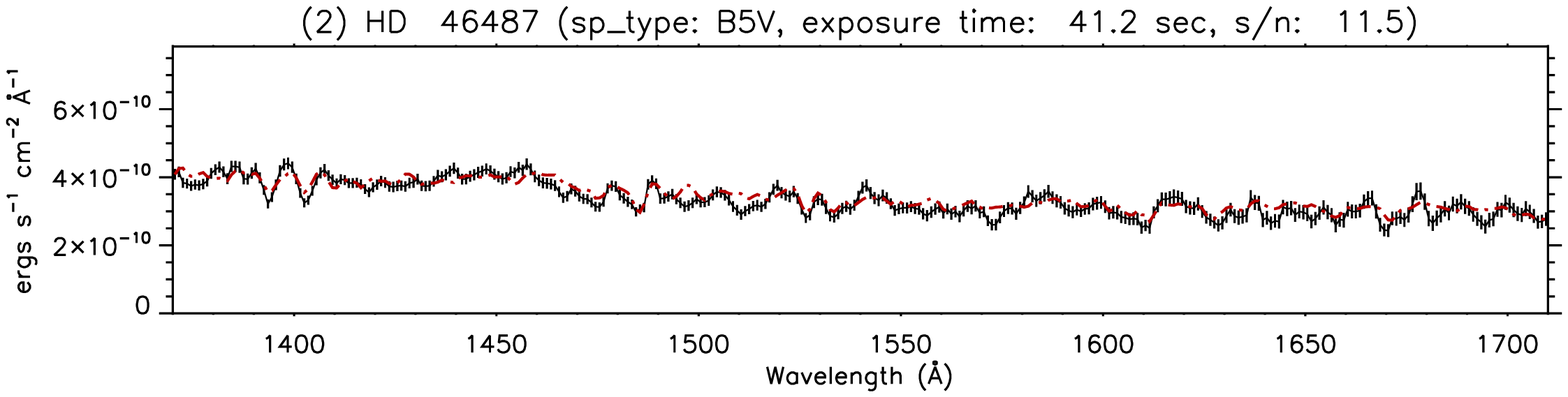}\\
  \includegraphics[width=13.4cm]{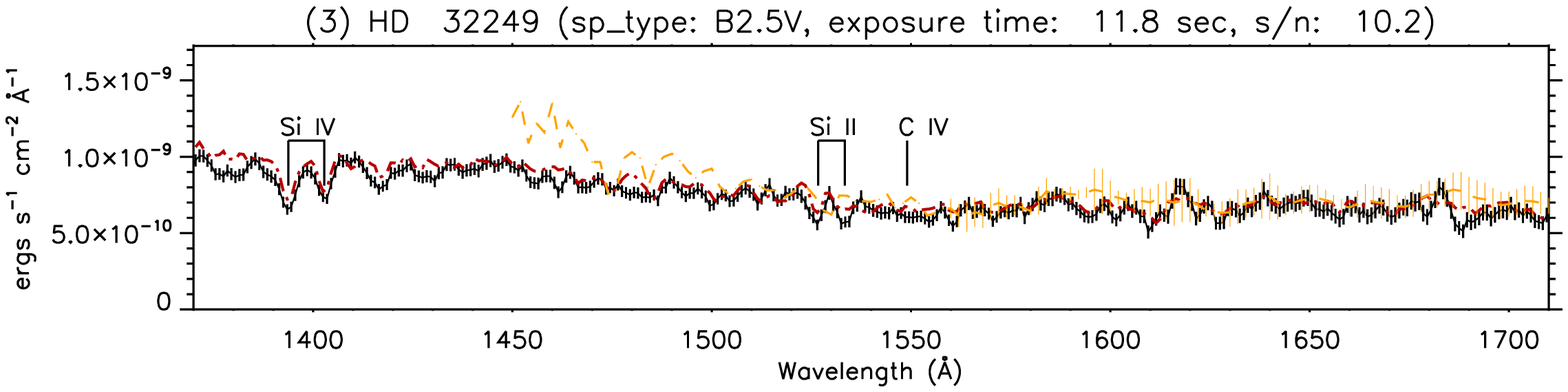}\\
  \includegraphics[width=13.4cm]{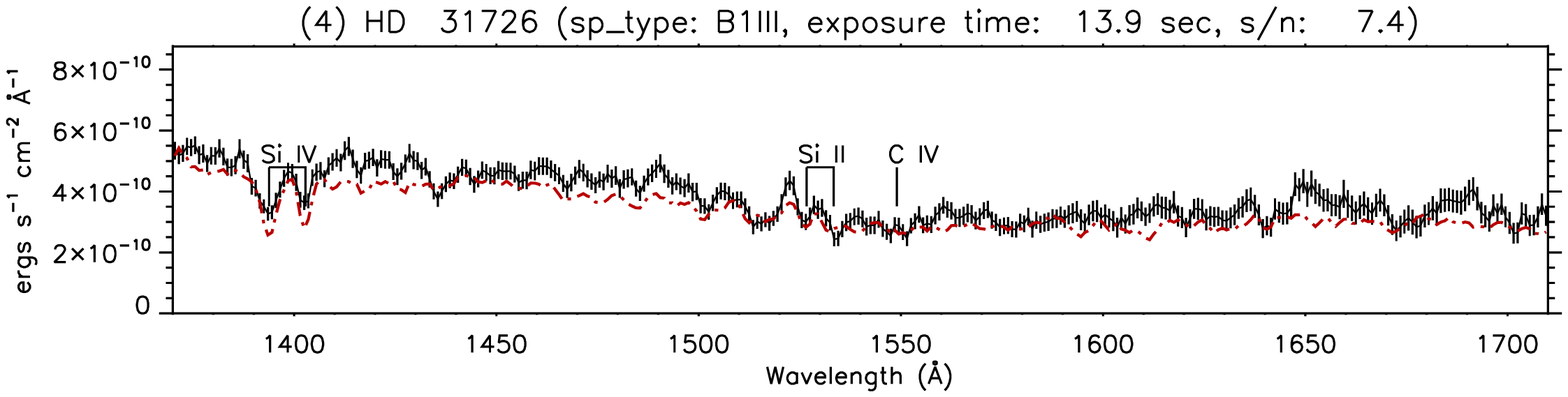}\\
  \includegraphics[width=13.4cm]{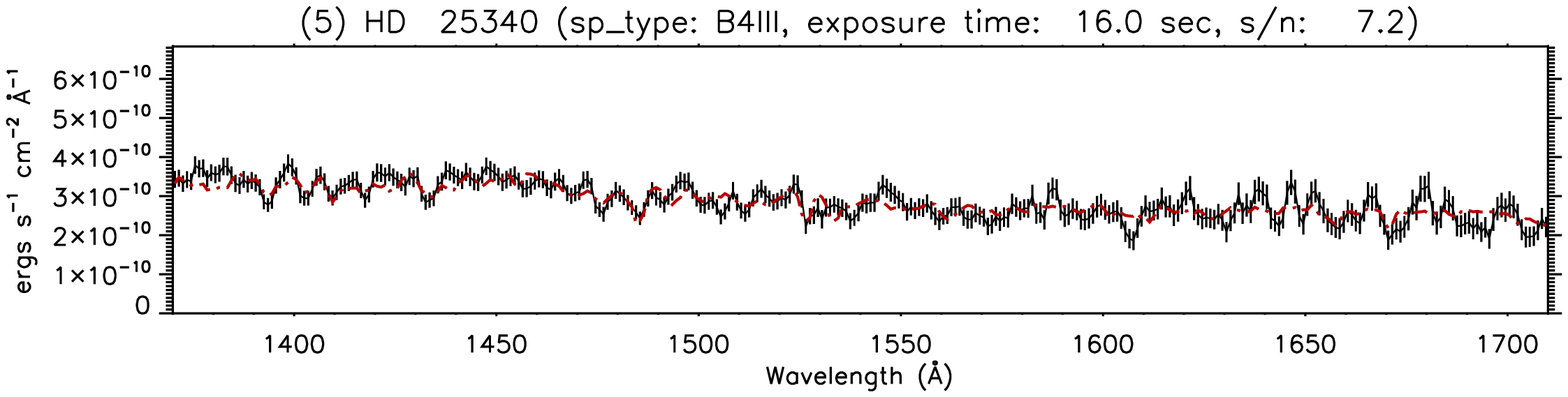}\\
  \includegraphics[width=13.4cm]{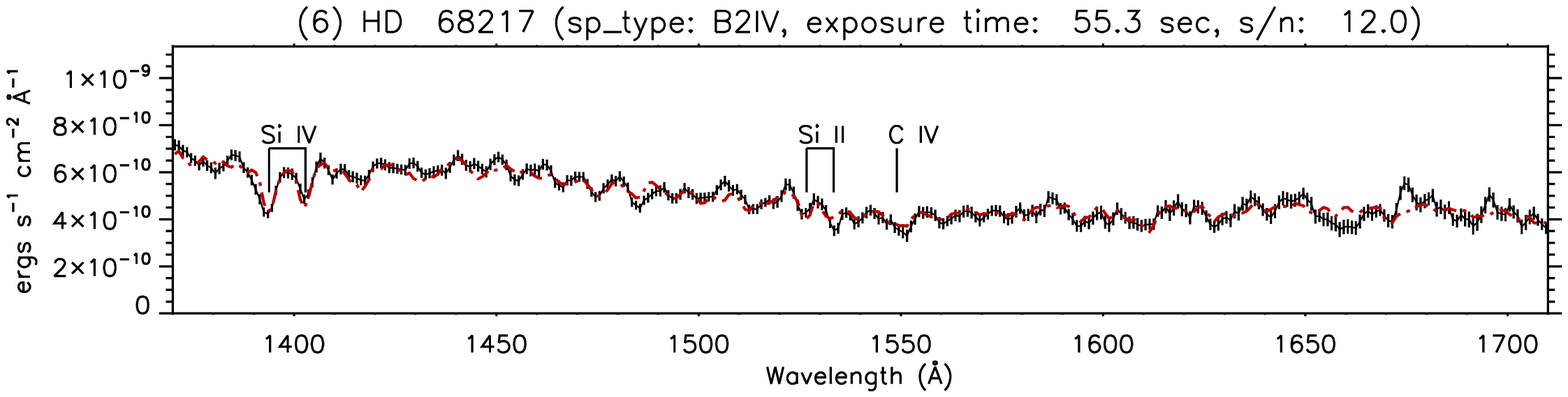}\\
 \end{center}
 \caption{Comparison of the \textit{FIMS} spectra with the \textit{IUE} spectra for
 the 28 reference stars listed in Table \ref{tbl:eff}. The solid black lines,
 dash-dotted red lines, and the dashed orange lines
 indicate the spectra observed by \textit{FIMS}, \textit{IUE}, and
 \textit{SKYLAB}, respectively.
 \label{fig:spectra}}
\end{figure*}

\addtocounter{figure}{-1}
\begin{figure*}
 \begin{center}
  \includegraphics[width=13.4cm]{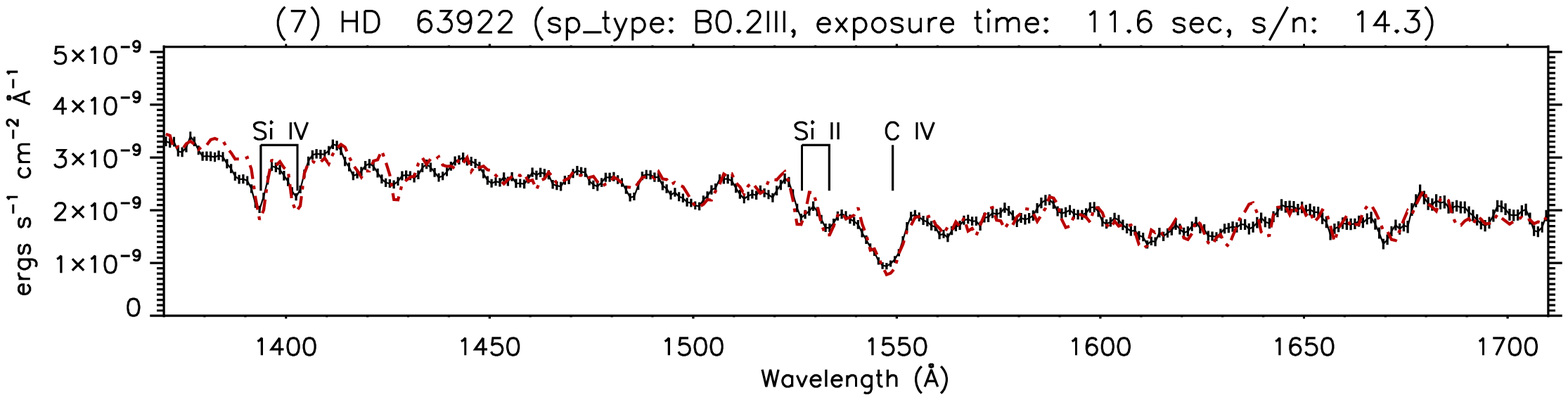}\\
  \includegraphics[width=13.4cm]{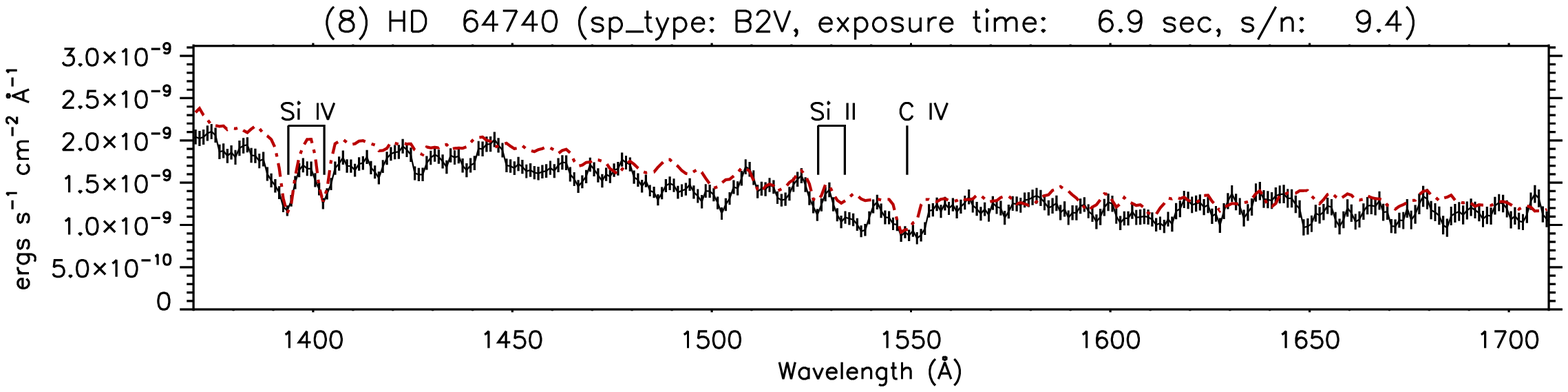}\\
  \includegraphics[width=13.4cm]{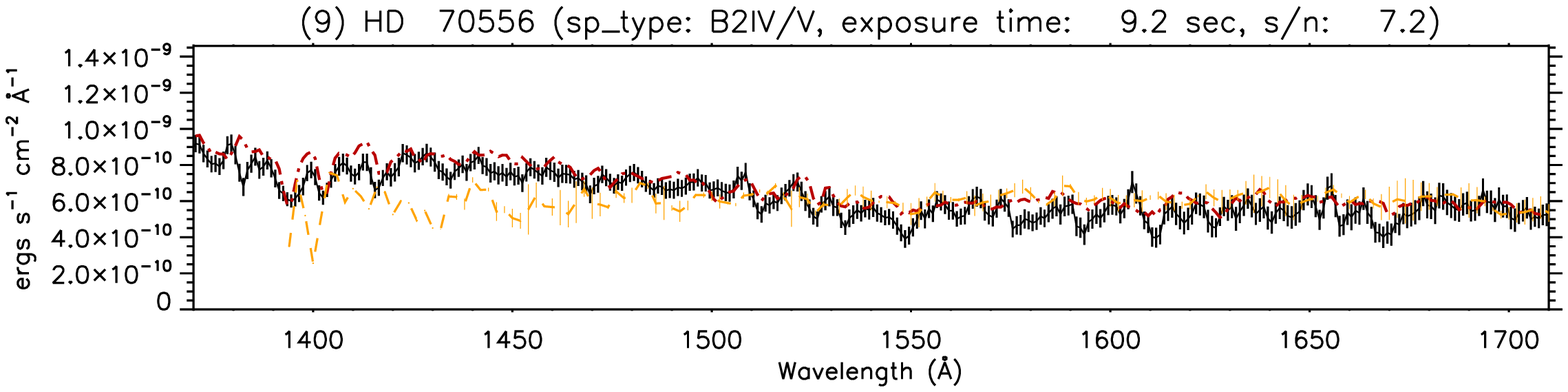}\\
  \includegraphics[width=13.4cm]{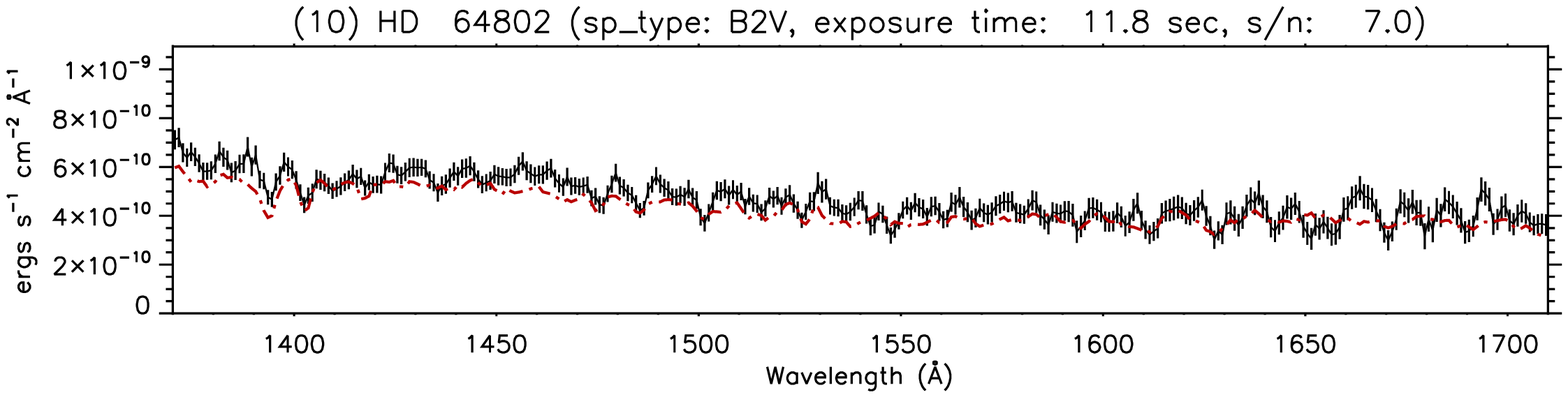}\\
  \includegraphics[width=13.4cm]{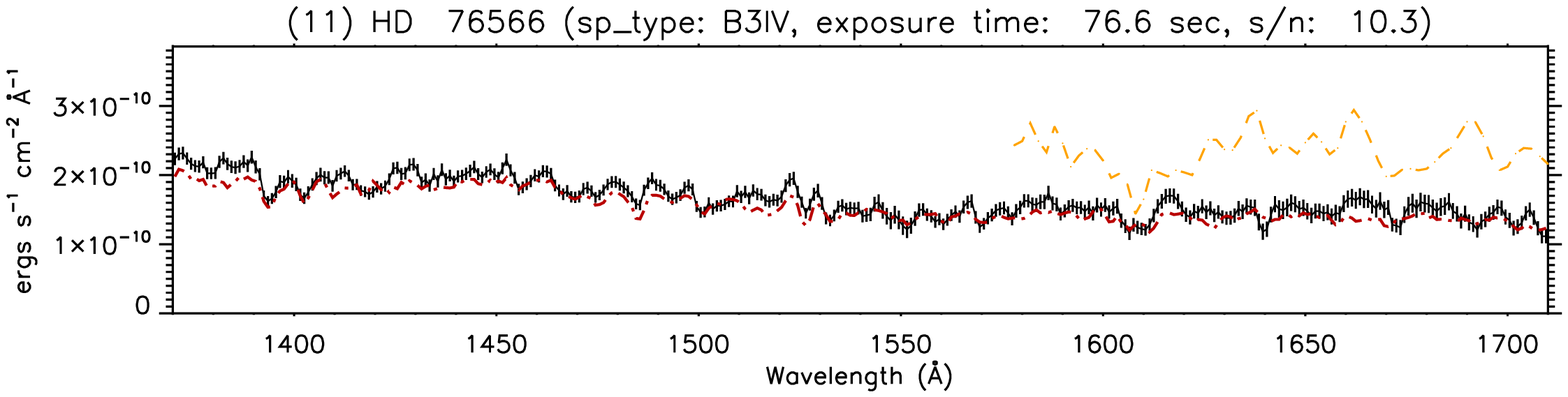}\\
  \includegraphics[width=13.4cm]{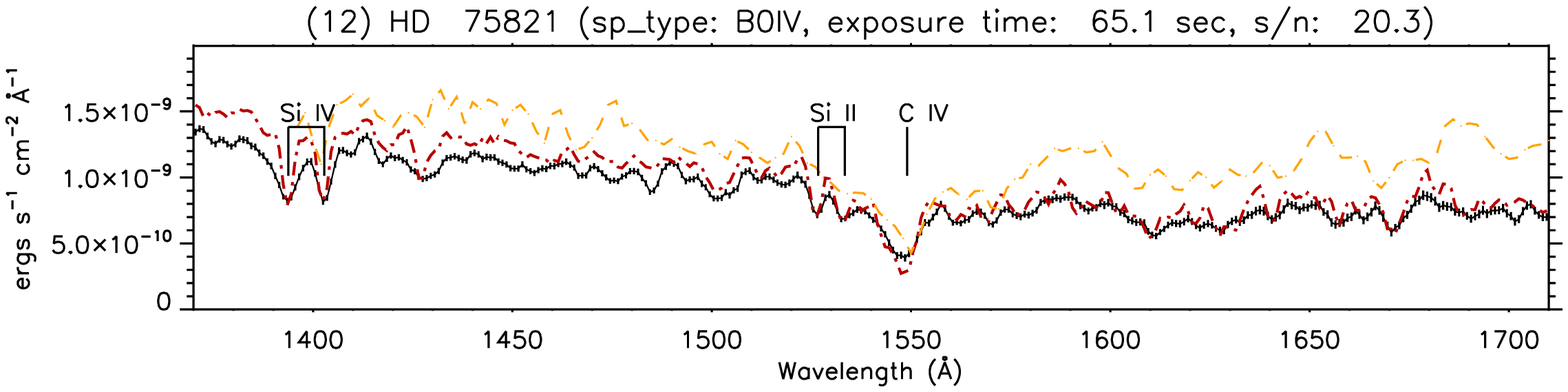}\\
  \includegraphics[width=13.4cm]{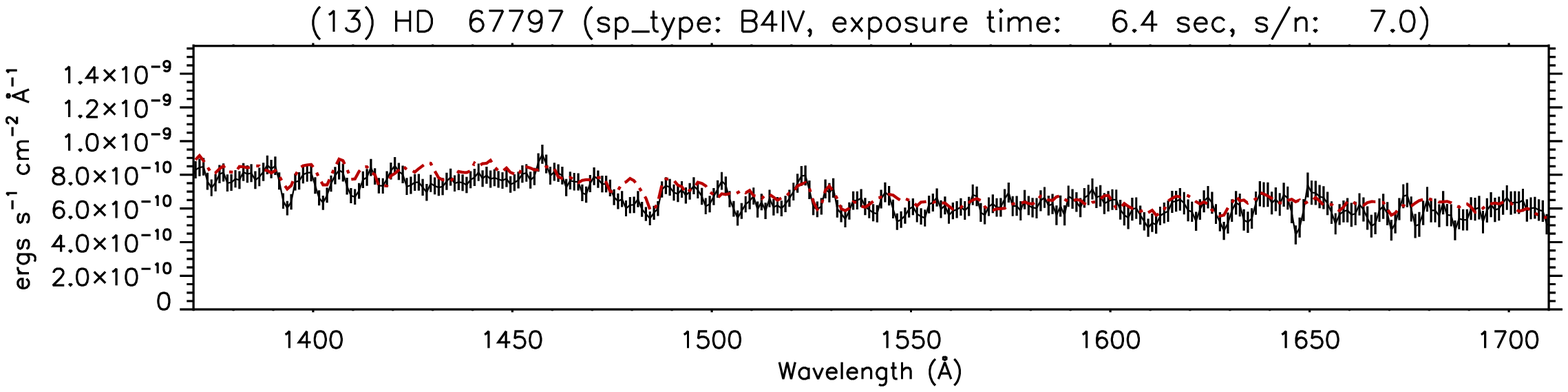}\\
 \end{center}
 \caption{(cont.)}
\end{figure*}

\addtocounter{figure}{-1}
\begin{figure*}
 \begin{center}
  \includegraphics[width=13.4cm]{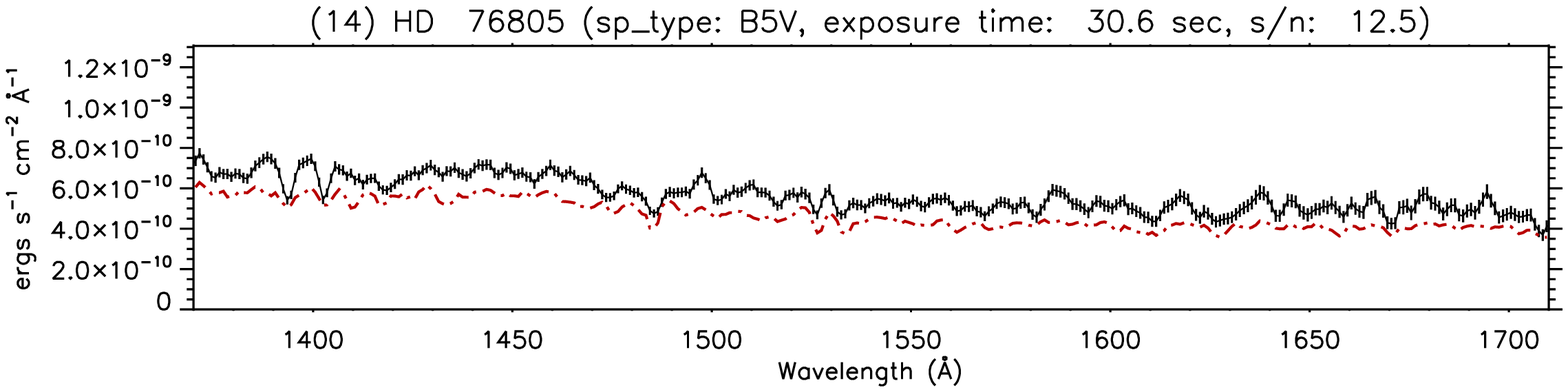}\\
  \includegraphics[width=13.4cm]{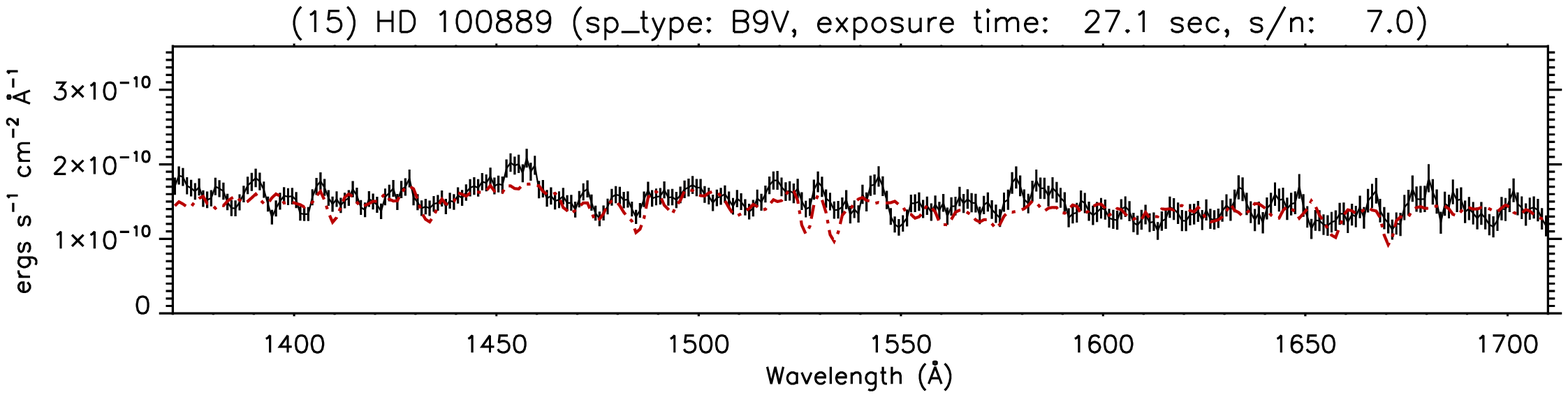}\\
  \includegraphics[width=13.4cm]{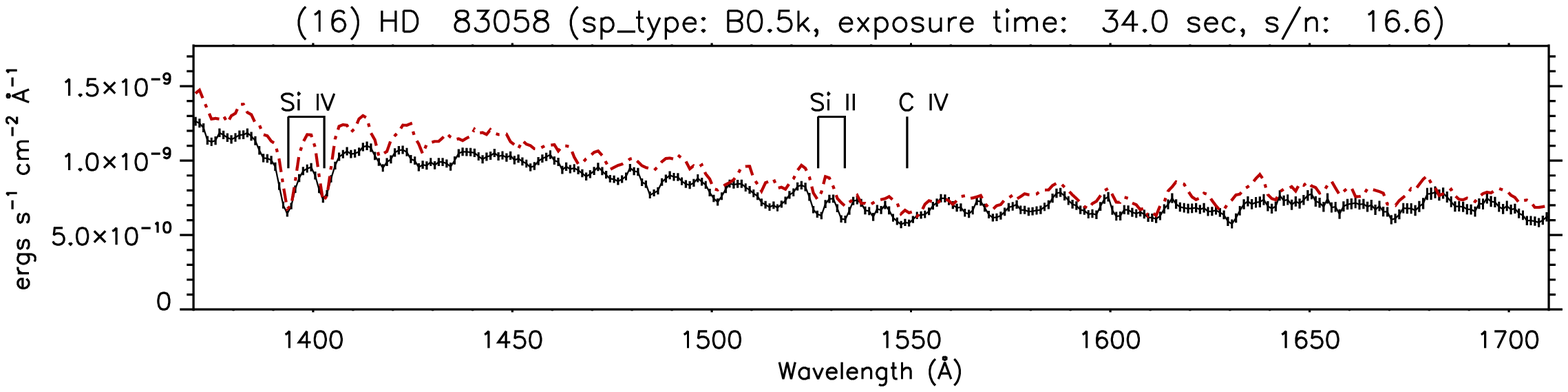}\\
  \includegraphics[width=13.4cm]{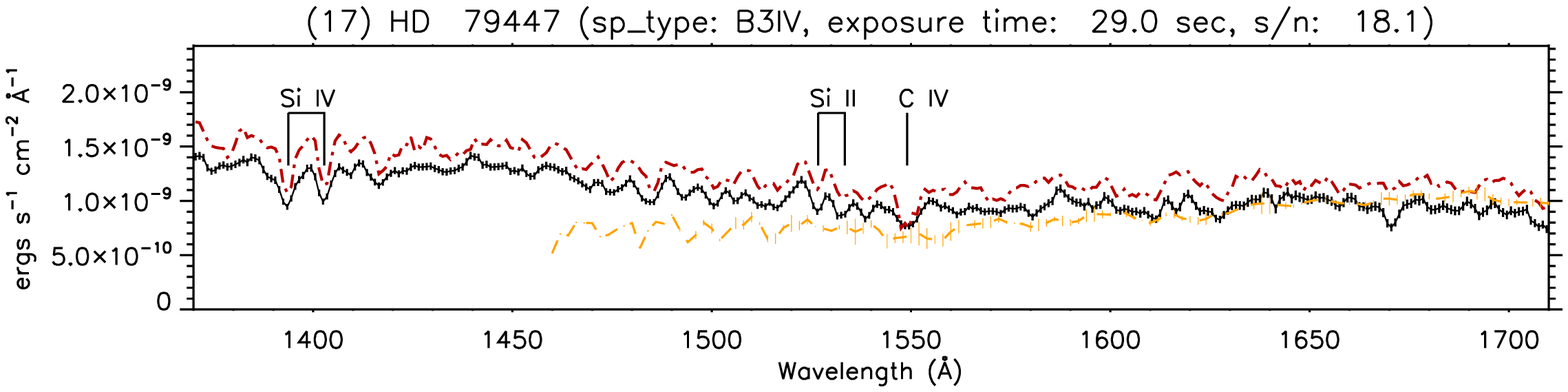}\\
  \includegraphics[width=13.4cm]{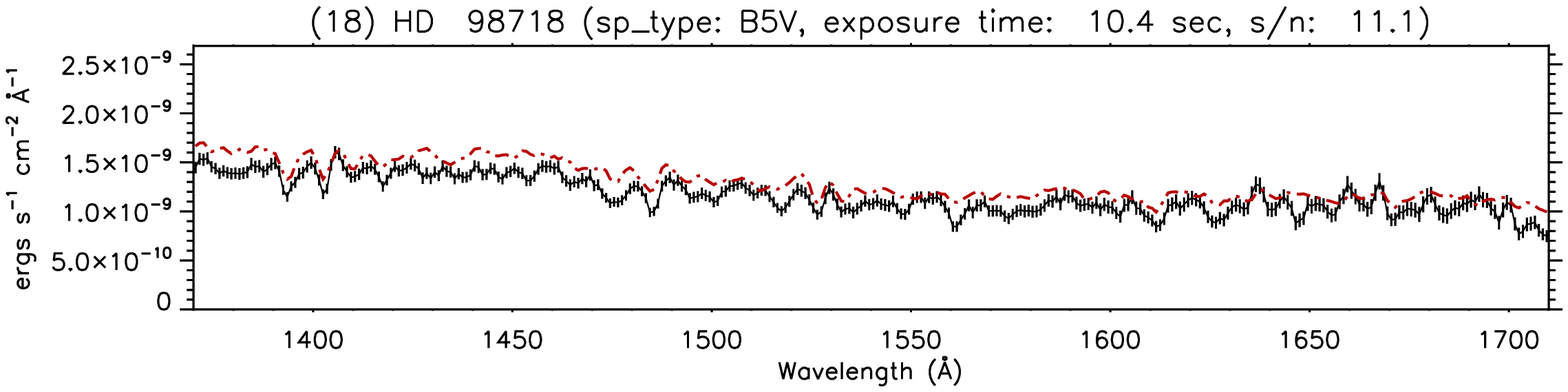}\\
  \includegraphics[width=13.4cm]{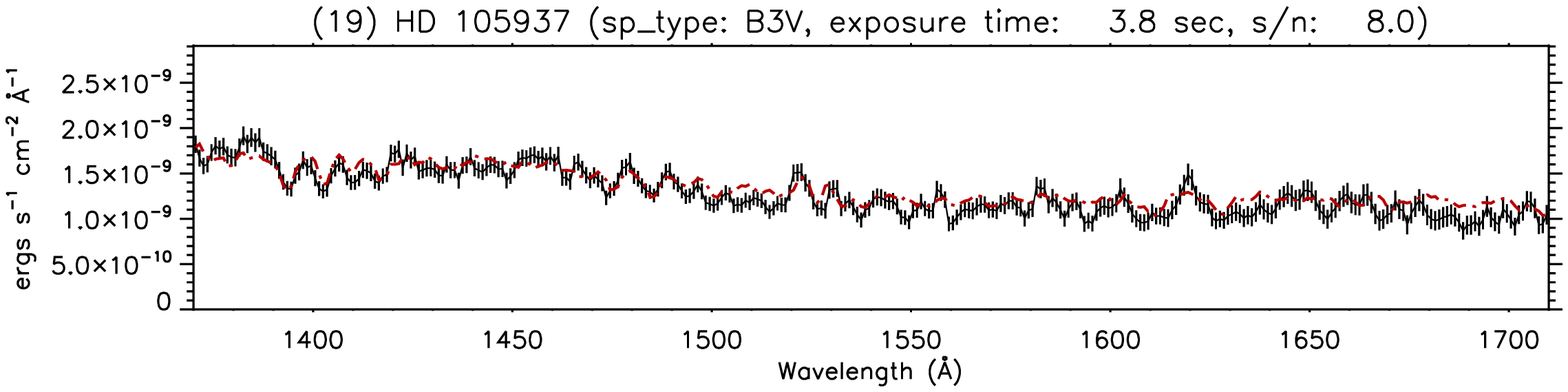}\\
  \includegraphics[width=13.4cm]{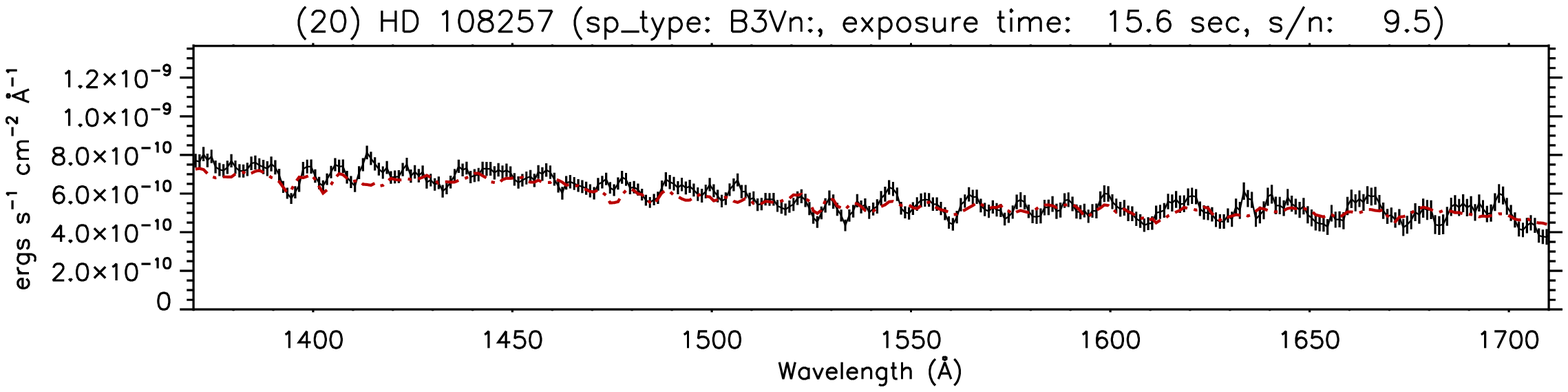}\\
 \end{center}
 \caption{(cont.)}
\end{figure*}

\addtocounter{figure}{-1}
\begin{figure*}
 \begin{center}
  \includegraphics[width=13.4cm]{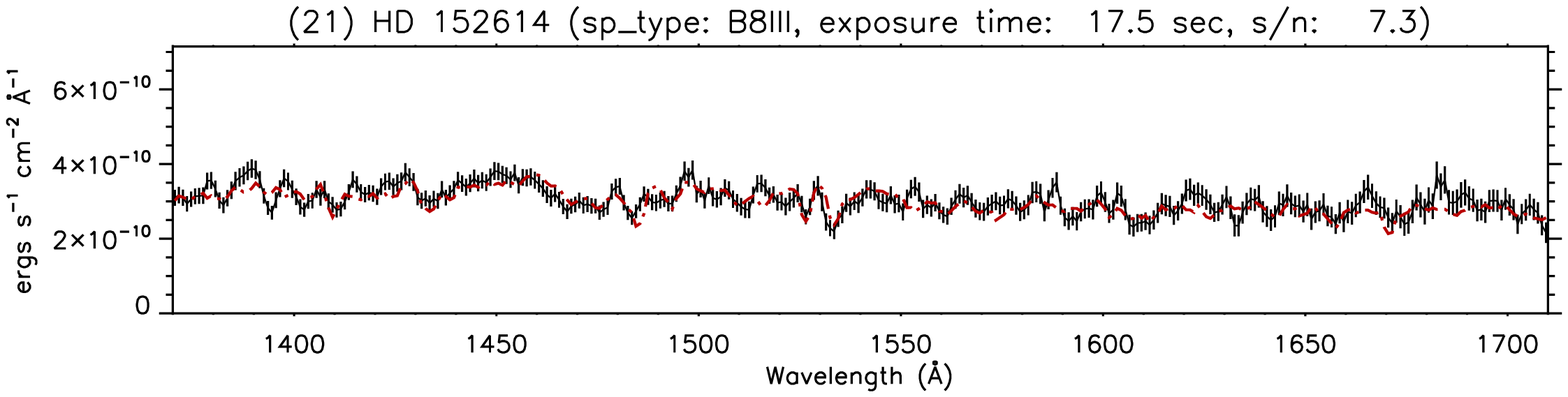}\\
  \includegraphics[width=13.4cm]{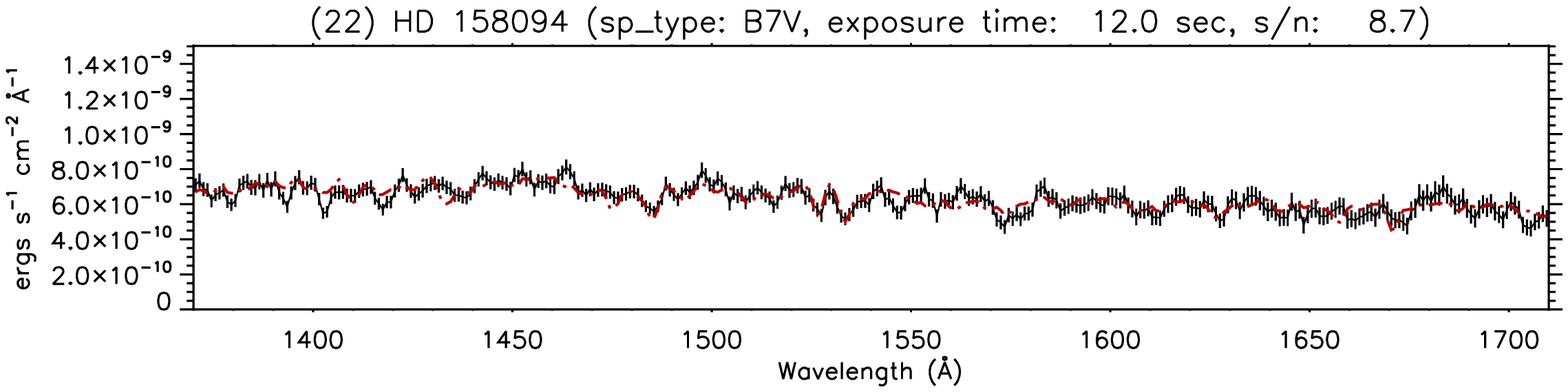}\\
  \includegraphics[width=13.4cm]{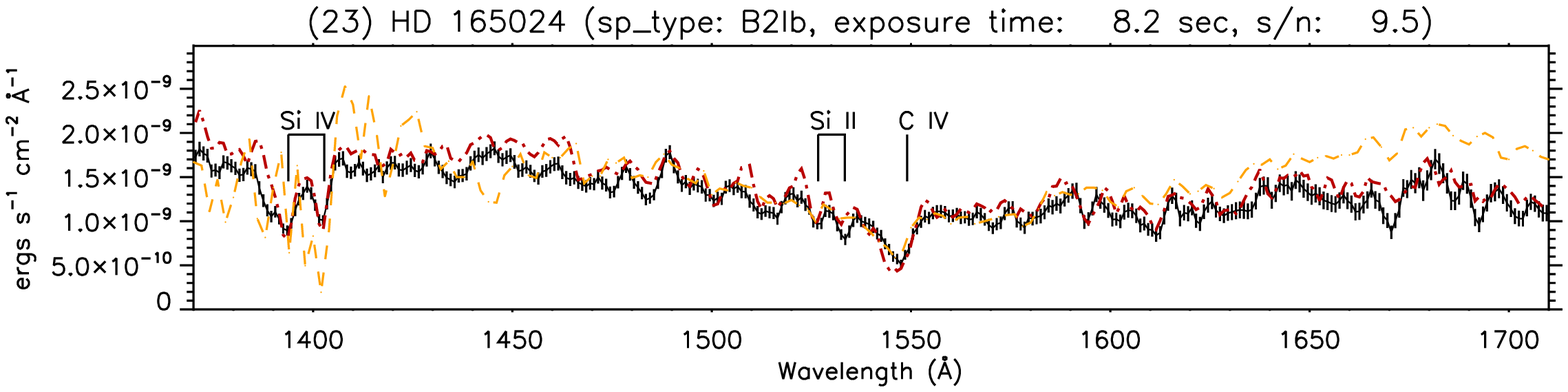}\\
  \includegraphics[width=13.4cm]{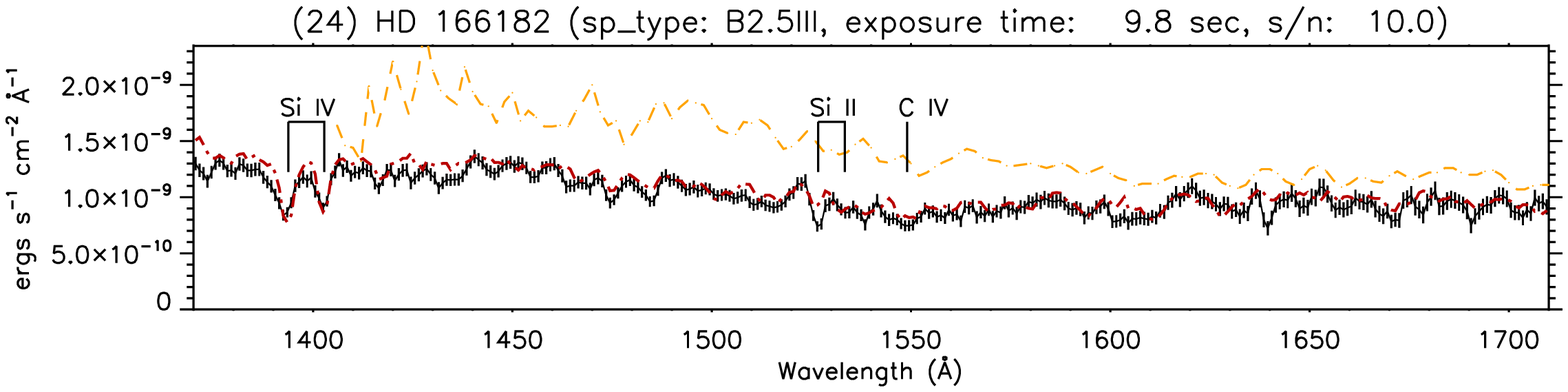}\\
  \includegraphics[width=13.4cm]{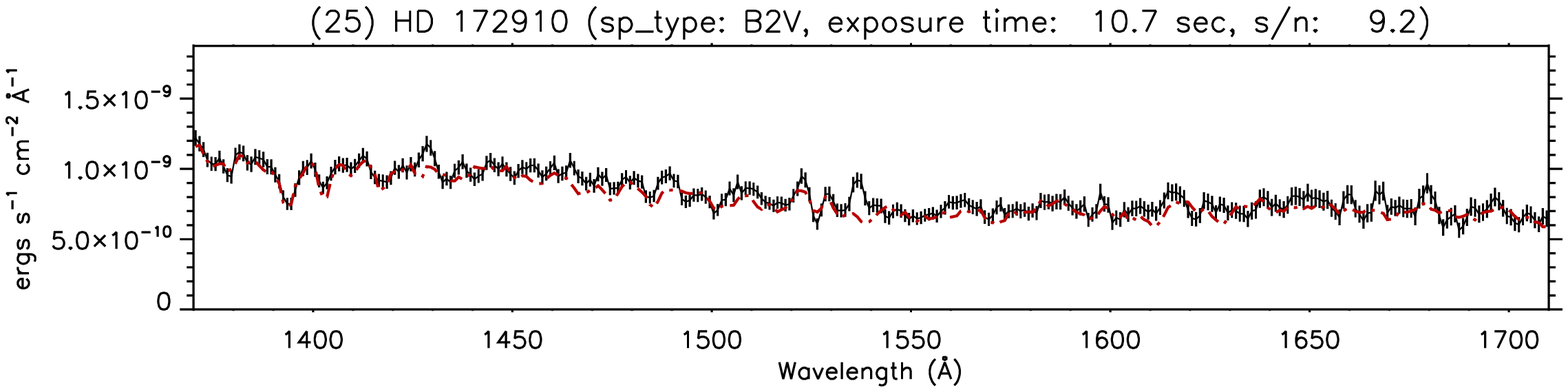}\\
  \includegraphics[width=13.4cm]{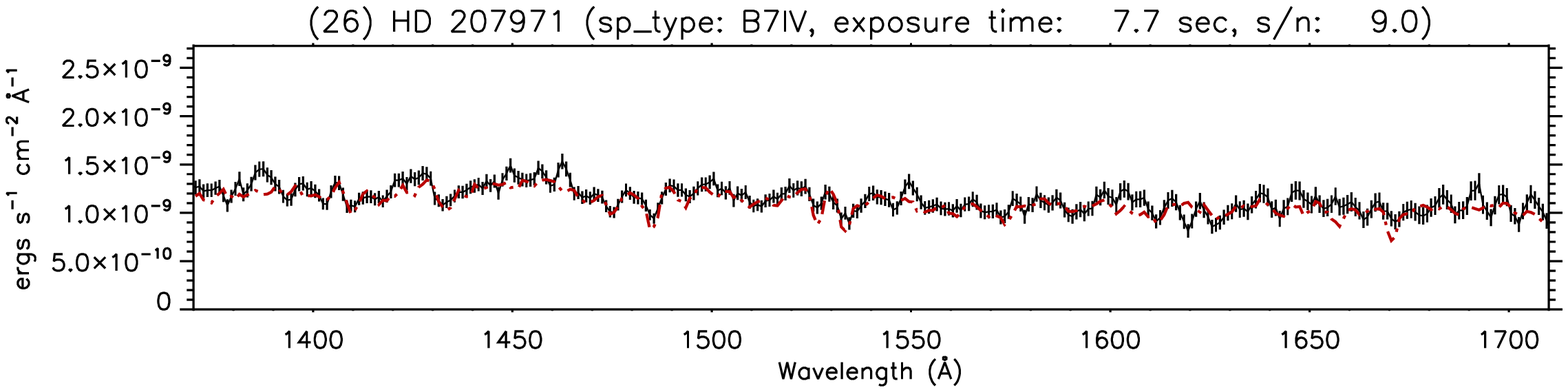}\\
  \includegraphics[width=13.4cm]{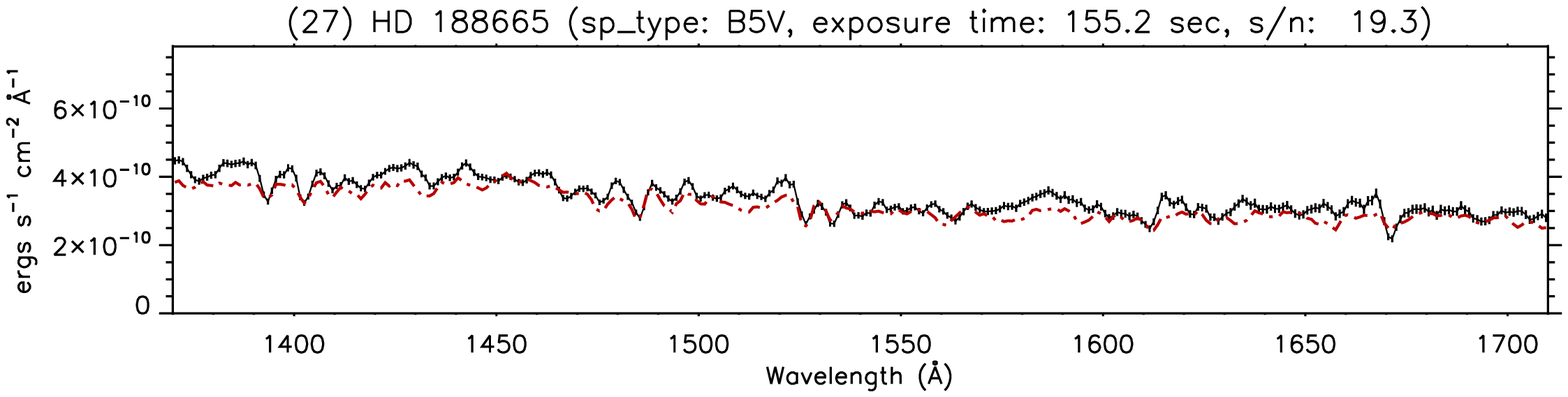}\\
 \end{center}
 \caption{(cont.)}
\end{figure*}

\addtocounter{figure}{-1}
\begin{figure*}
 \begin{center}
  \includegraphics[width=13.4cm]{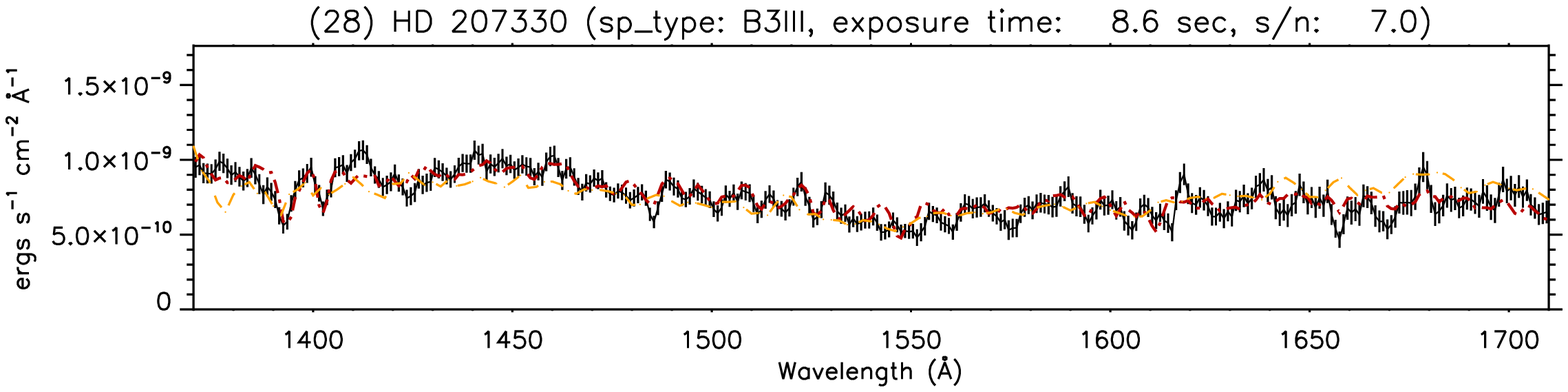}\\
 \end{center}
 \caption{(cont.)}
\end{figure*}

\begin{figure*}
 \begin{center}
  \includegraphics[width=8cm]{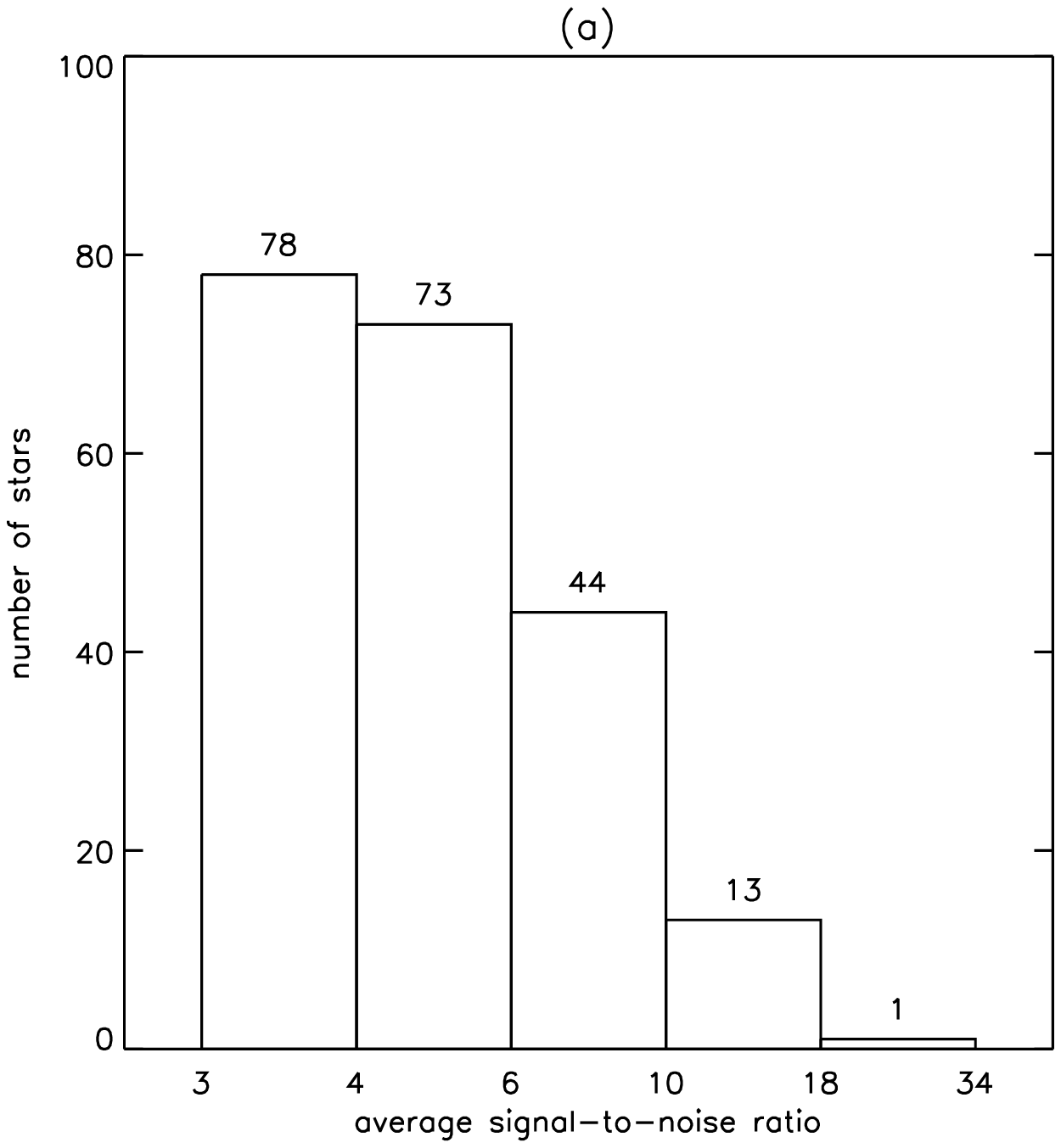}\quad
  \includegraphics[width=8cm]{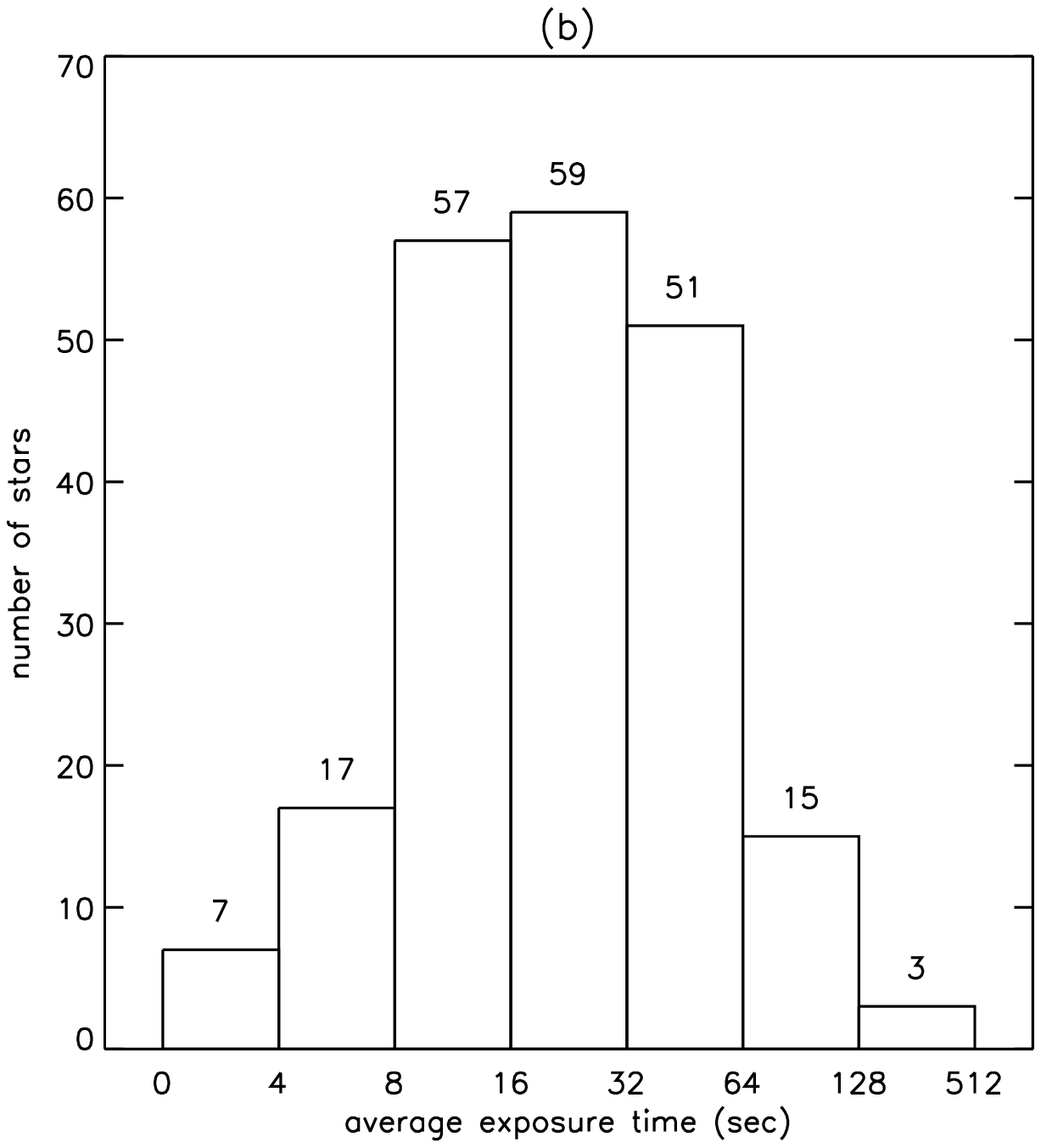}\\
  \includegraphics[width=8cm]{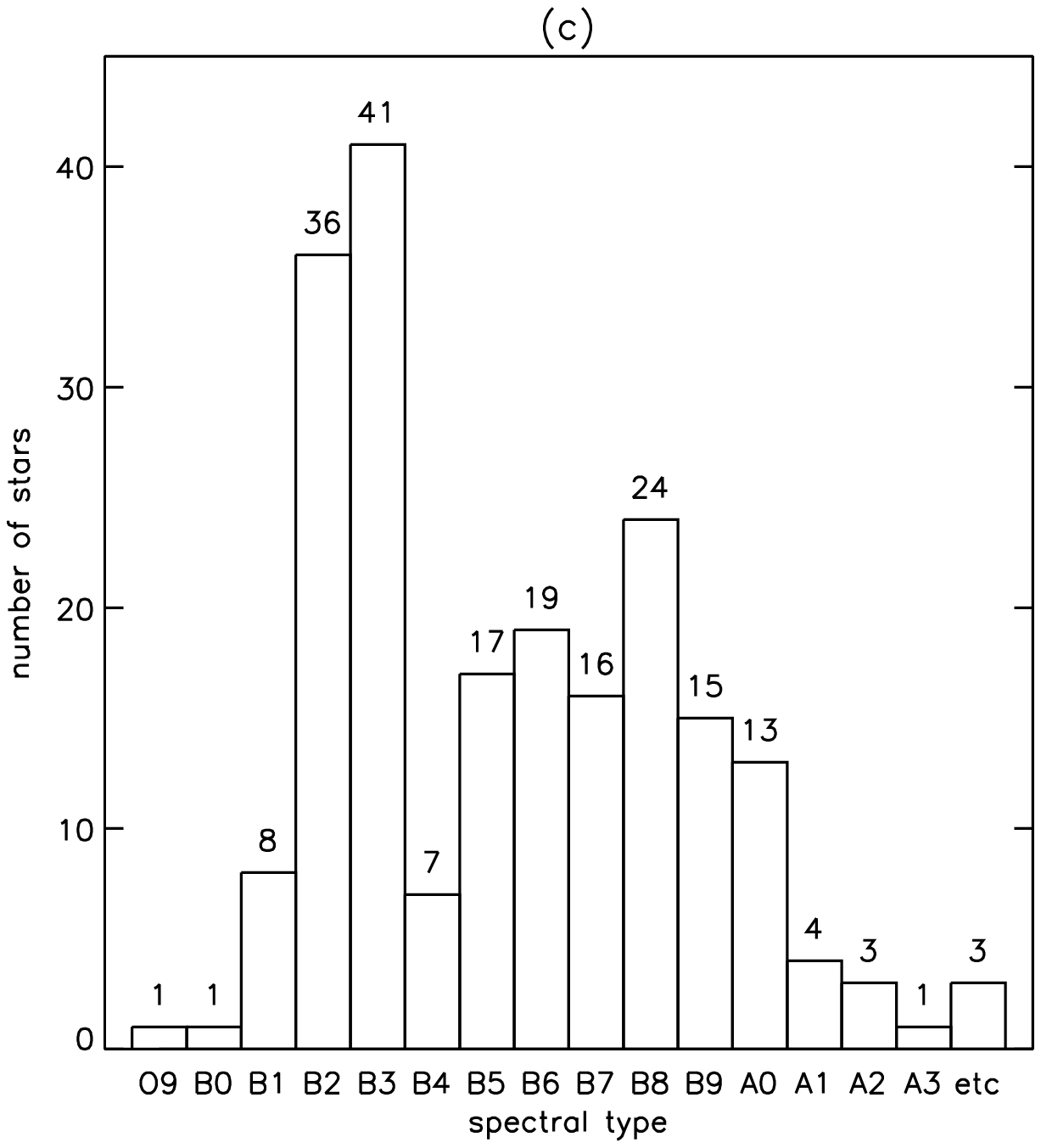}\quad
  \includegraphics[width=8cm]{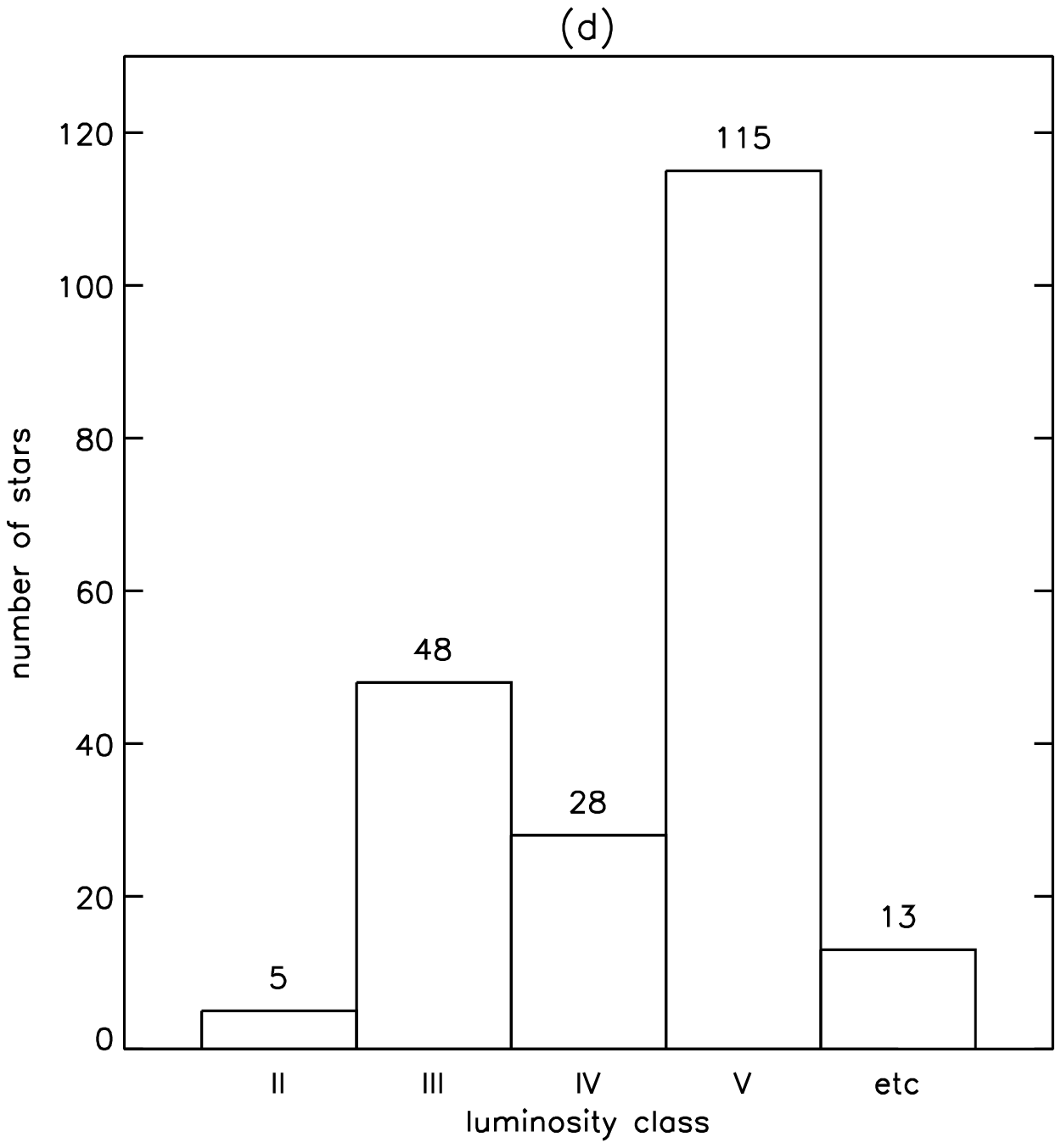}\\
 \end{center}
 \caption{Statistics for the 209  \textit{FIMS}
 catalogue stars indicating the number of stars against (a) the average
 SNR per angstrom, (b) exposure time, (c) spectral type, and (d)
 luminosity class. \label{fig:histo}}
\end{figure*}

\begin{table*}
 \centering
  \begin{minipage}{18cm}
  \caption{List of sample stars whose spectra are presented in Figure \ref{fig:sample}. \label{tbl:sample}}
  \begin{tabular}{@{}ccccccccccc@{}}
  \hline
  HD & R.A. & Decl. & \textit{l} & \textit{b} %
  & FIMS & FIMS & FIMS & SpType & SpType ref & Comment \\
  ID & (deg) & (deg) & (deg) & (deg) %
  & Flux\footnote{units: erg s$^{-1}$ cm$^{-2}$ {\AA}$^{-1}$} & S/N & Exp\footnote{units: sec} %
  & \citep{ski14} & \citep{ski14} & \\
  (1) & (2) & (3) & (4) & (5) & (6) & (7) & (8) & (9) & (10) & (11) \\
  \hline
  76728 & 133.8 & -60.6 & 277.7 & -10.0 & 5.15E-10 & 15.0 & 37.6 & B8II        & \citet{gar94}  & UVSST $\&$ SKYLAB \\
  69302 & 123.6 & -45.8 & 262.0 & -6.2  & 3.17E-10 & 14.2 & 76.5 & B2IV        & \citet{hou78}  & UVSST $\&$ SKYLAB \\
  70839 & 125.3 & -58.0 & 272.9 & -11.9 & 2.04E-10 & 8.0  & 37.3 & B2V         & \citet{cuc76}  & UVSST \\
  83944 & 144.8 & -61.3 & 281.9 & -6.6  & 2.10E-10 & 7.4  & 29.6 & B9V         & \citet{zor09}  & SKYLAB \\
  84567 & 146.3 & -30.2 & 261.8 & 17.5  & 1.38E-10 & 5.1  & 19.2 & B0IV        & \citet{hou82}  & FIMS only \\
  65176 & 119.4 & -1.6  & 222.1 & 13.9  & 3.29E-11 & 3.4  & 33.1 & B1.5Ib/IIep & \citet{sch67}  & FIMS only \\
  \hline
  \end{tabular}
  \end{minipage}
\end{table*}

\begin{figure*}
 \begin{center}
  \includegraphics[width=13.4cm]{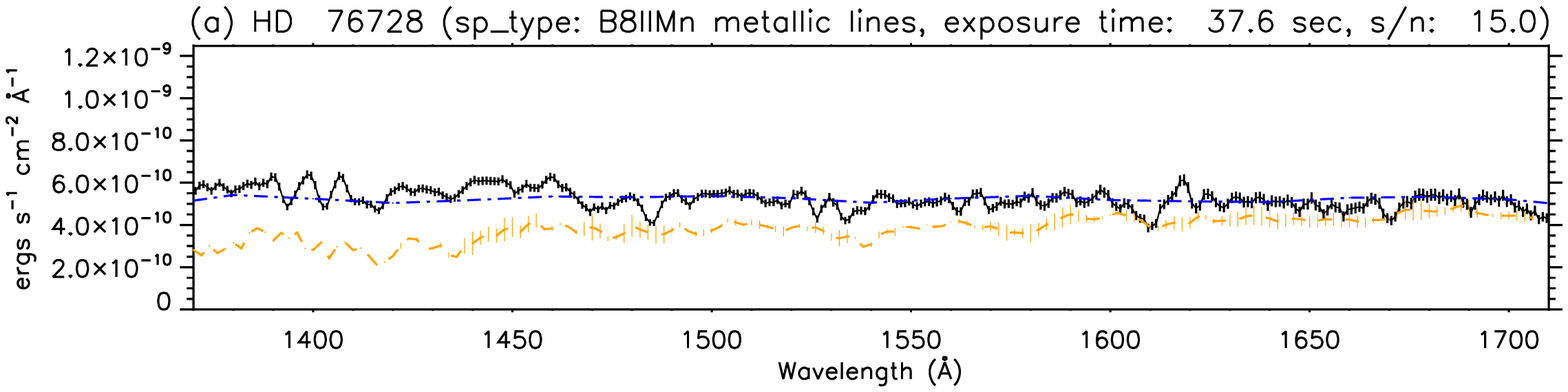}\\\vspace{10pt}
  \includegraphics[width=13.4cm]{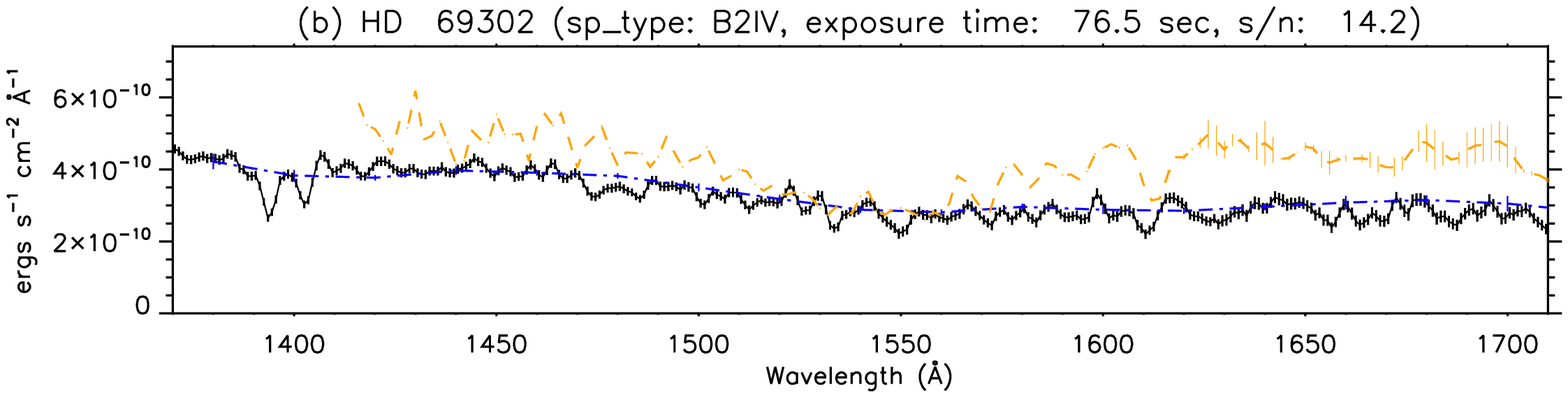}\\\vspace{10pt}
  \includegraphics[width=13.4cm]{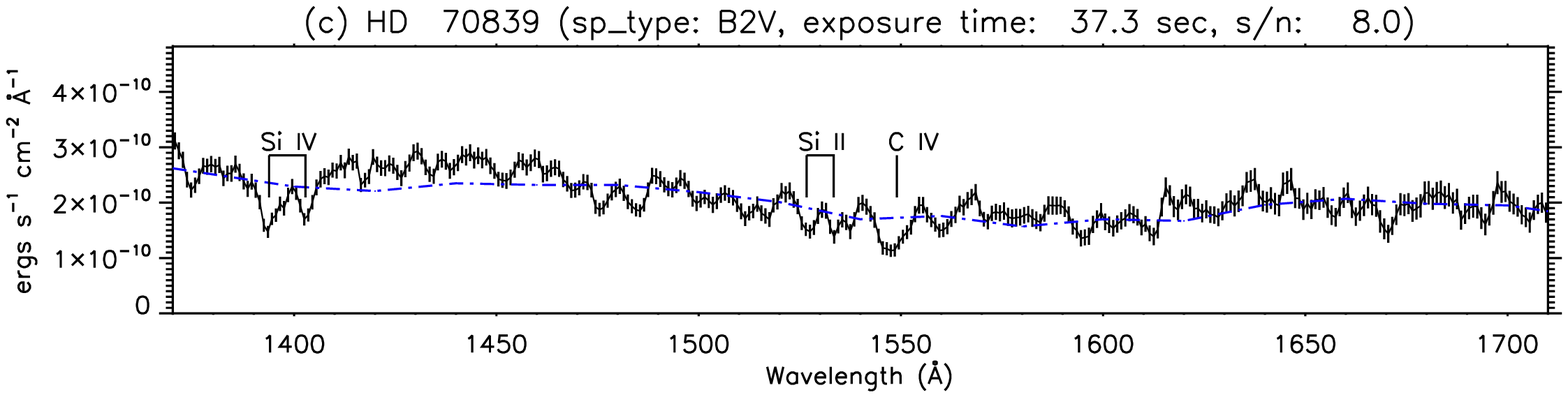}\\\vspace{10pt}
  \includegraphics[width=13.4cm]{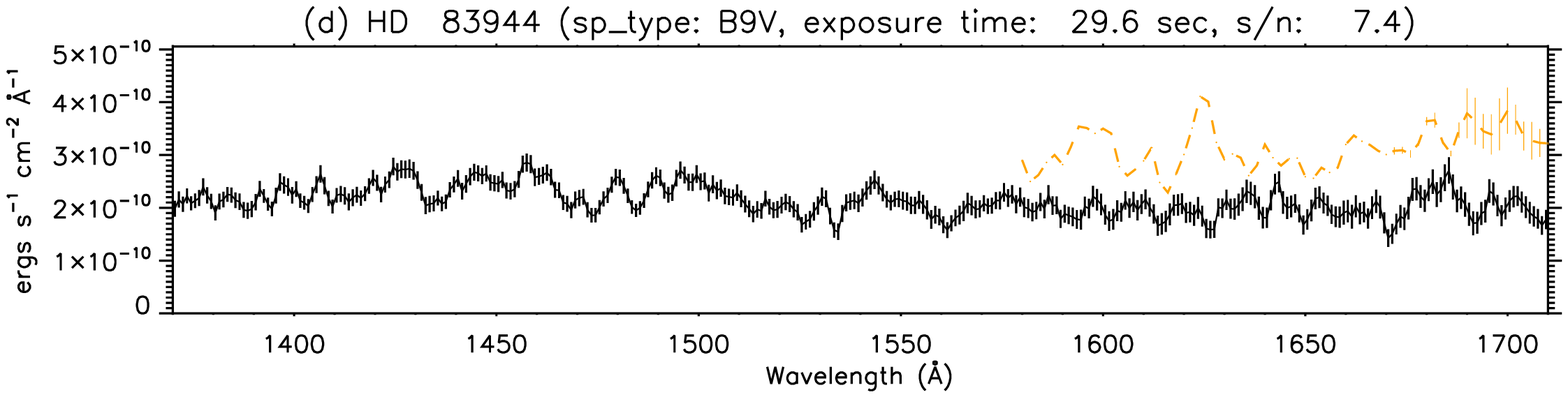}\\\vspace{10pt}
  \includegraphics[width=13.4cm]{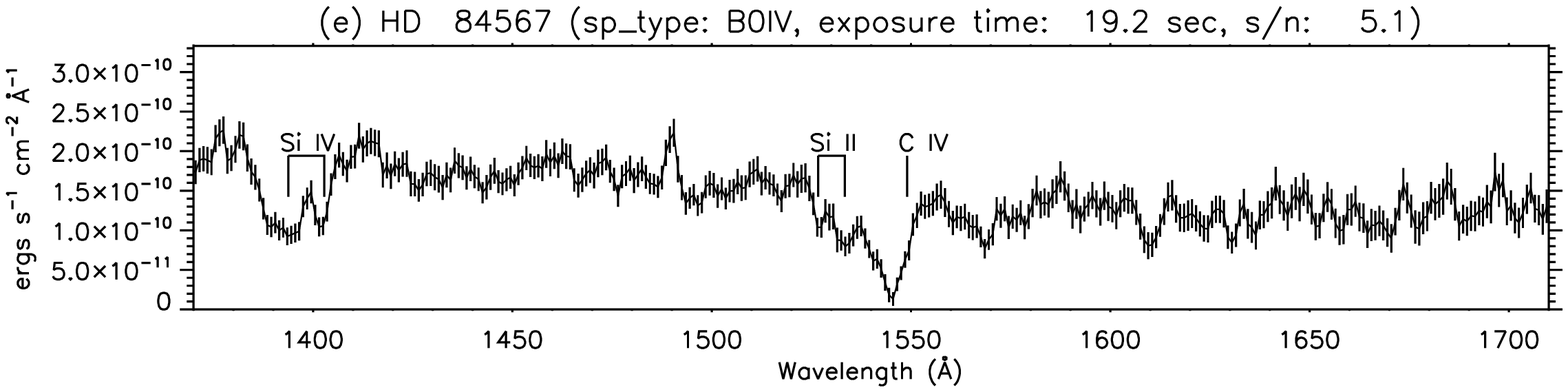}\\\vspace{10pt}
  \includegraphics[width=13.4cm]{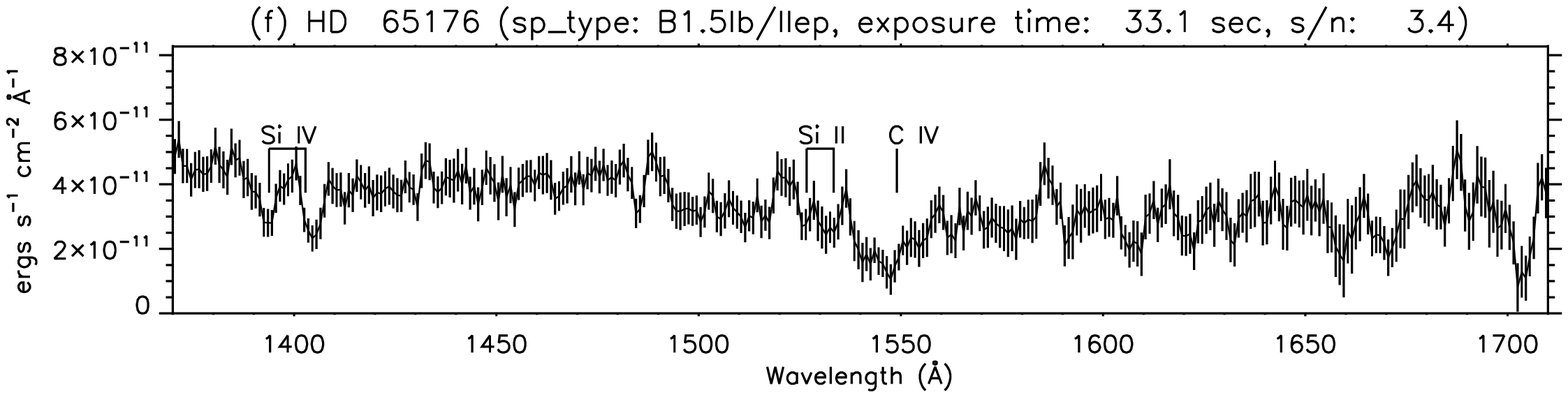}\\
 \end{center}
 \caption{(a)-(f)  Spectra of the sample stars listed in Table \ref{tbl:sample}.
 The solid black line, dash-dotted blue line, and dashed orange
 line  indicate the spectra observed by the \textit{FIMS}, \textit{UVSST},
 and \textit{SKYLAB}, respectively.
 \label{fig:sample}}
\end{figure*}

\begin{figure*}
 \begin{center}
  \includegraphics[width=13.4cm]{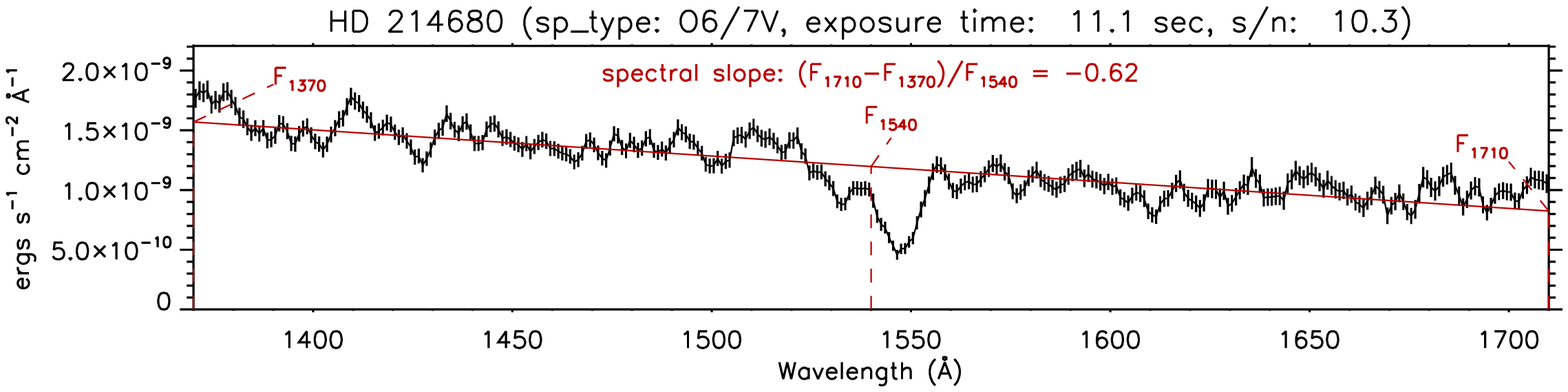}\\
 \end{center}
 \caption{Linear fit of the spectrum of HD 214680.\label{fig:slope_ex}}
\end{figure*}

\section{Results and Discussion}

We present the statistics and several representative spectra of the
bright stars catalogued by the \textit{FIMS} in this section. In
Figure \ref{fig:flux}, the fluxes of the \textit{FIMS},
\textit{TD1}-F1565, and \textit{SKYLAB} are compared with those of
the \textit{IUE}; all are averaged in the same manner over similar
wavelength ranges for 323 stars observed by the \textit{IUE} with a
large aperture mode. More than 92$\%$ of the \textit{FIMS} fluxes,
denoted by black plus (+) symbols, agree well with those of the
\textit{IUE} within the \textit{FIMS} systemic error of 25$\%$
\citep{ede06b}. The \textit{TD1}-F1565 fluxes, marked with red
asterisk symbols, also agree well with those of \textit{IUE} within
a 25$\%$ error for 96$\%$ of the 323 stars. However, large
fluctuations are seen in the \textit{SKYLAB} fluxes, marked with
blue cross ($\times$) symbols; only 65$\%$ of 86 stars observed by
\textit{SKYLAB} show fluxes comparable to those of the \textit{IUE}
within a 25$\%$ error.

In order to verify the \textit{FIMS} spectra after the correction of
effective area, they are compared in Figure \ref{fig:spectra} with
the \textit{IUE} spectra for the 28 stars listed in Table
\ref{tbl:eff}, together with the available \textit{SKYLAB} spectra.
The black solid lines, dash-dotted red lines, and dashed orange
lines indicate the \textit{FIMS}, \textit{IUE}, and \textit{SKYLAB}
spectra, respectively. The \textit{FIMS} spectra are seen to match
well with the corresponding \textit{IUE} spectra. For some stars,
such as (14) HD 76805, (16) HD 83058, (17) HD 79447, and (18) HD
98718, the flux levels between the \textit{FIMS} and the
\textit{IUE} observations show discrepancies due to the fluctuation
in the effective area but they are all within 25$\%$ of the
\textit{FIMS} systematic error range \citep{ede06b}. The prominent
absorption lines of early-type stars, such as \mbox{C\,{\sc iv}}
$\lambda\lambda$1548, 1551 and \mbox{Si\,{\sc iv}}
$\lambda\lambda$1394, 1403 features, are clearly seen in the
\textit{FIMS} spectra. However, narrower lines, such as
\mbox{Si\,{\sc ii$^*$}} $\lambda\lambda$1527, 1533, are less clear
in the \textit{FIMS} spectra than in the \textit{IUE} spectra
because of the variation of spectral resolution across the detector.
The \textit{SKYLAB} spectra show large deviation from the
corresponding \textit{IUE} and \textit{FIMS} spectra. This
demonstrates the superior quality of \textit{FIMS} data compared to
those of {\em SKYLAB}.

Figure \ref{fig:histo} presents statistics for the 209 stars of the
\textit{FIMS} catalogue, excluding those observed by the
\textit{IUE}. Figures \ref{fig:histo}(a) to \ref{fig:histo}(d) are
histograms of the average SNR per angstrom, exposure time, spectral
type, and luminosity class, respectively. The figures illustrate
that, of the 209  catalogue stars, the spectra of 58 stars had high
SNRs above 6.0. Most of these were bright stars observed with longer
exposure times than the average of 20 s. The stellar spectral type
of catalogue stars ranged from O9  to A3; 115 of these stars are
main sequence stars, 28 are subgiant stars, 48 are normal giant
stars, 5 are bright giant stars, and 13 are unknown. The stars with
spectral types beyond A4 were not analysed with a sufficient SNR
because their flux level was weak in the FUV band due to the low
surface temperature; hence, they have been excluded from the
catalogue.

For the stars listed in Table \ref{tbl:sample}, example spectra are
plotted in Figure \ref{fig:sample} in order to compare the
\textit{FIMS} spectra with the previous observations of
\textit{UVSST} and \textit{SKYLAB}. The example spectra were
arbitrarily chosen and displayed in decreasing order of SNR to show
variation of the spectral quality. The \textit{FIMS} spectra are
indicated using solid black lines in the figure. It is seen that the
\textit{UVSST} spectra, marked with dash-dotted blue lines in
Figures \ref{fig:sample}(a) to \ref{fig:sample}(c) agree well with
the \textit{FIMS} spectra; however, the absorption features are not
resolved  in the \textit{UVSST} spectra due to the low spectral
resolution of 35 {\AA}. The \textit{SKYLAB} spectra, which are
marked using dashed orange lines in Figures \ref{fig:sample}(a),
\ref{fig:sample}(b), and \ref{fig:sample}(d), have a similar
spectral resolution to that of the \textit{FIMS} spectra. However,
as seen in Figure \ref{fig:flux}, the flux levels of the
\textit{SKYLAB} spectra deviate significantly from those of the
\textit{IUE} spectra while the \textit{FIMS} flux levels are more or
less consistent with the \textit{IUE} observations. Hence, we
believe that the \textit{FIMS} spectra are more reliable than the
\textit{SKYLAB} spectra. Figures \ref{fig:sample}(e) and
\ref{fig:sample}(f) show two examples observed by the \textit{FIMS}
only. Strong absorption lines including \mbox{Si\,{\sc iv}},
\mbox{Si\,{\sc ii$^*$}}, and \mbox{C\,{\sc iv}} lines are seen in
both spectra even though the SNR is low in the case of Figure
\ref{fig:sample}(f). According to the SIMBAD database, the spectral
type of HD 65176, shown in Figure \ref{fig:sample}(f), is A0. It
should be noted that the corresponding spectrum of A0 is not
expected to have such prominent absorption features in the
\mbox{Si\,{\sc iv}} and \mbox{C\,{\sc iv}} lines. We will discuss
this discrepancy in the following text.

\begin{figure*}
 \begin{center}
  \includegraphics[width=13.4cm]{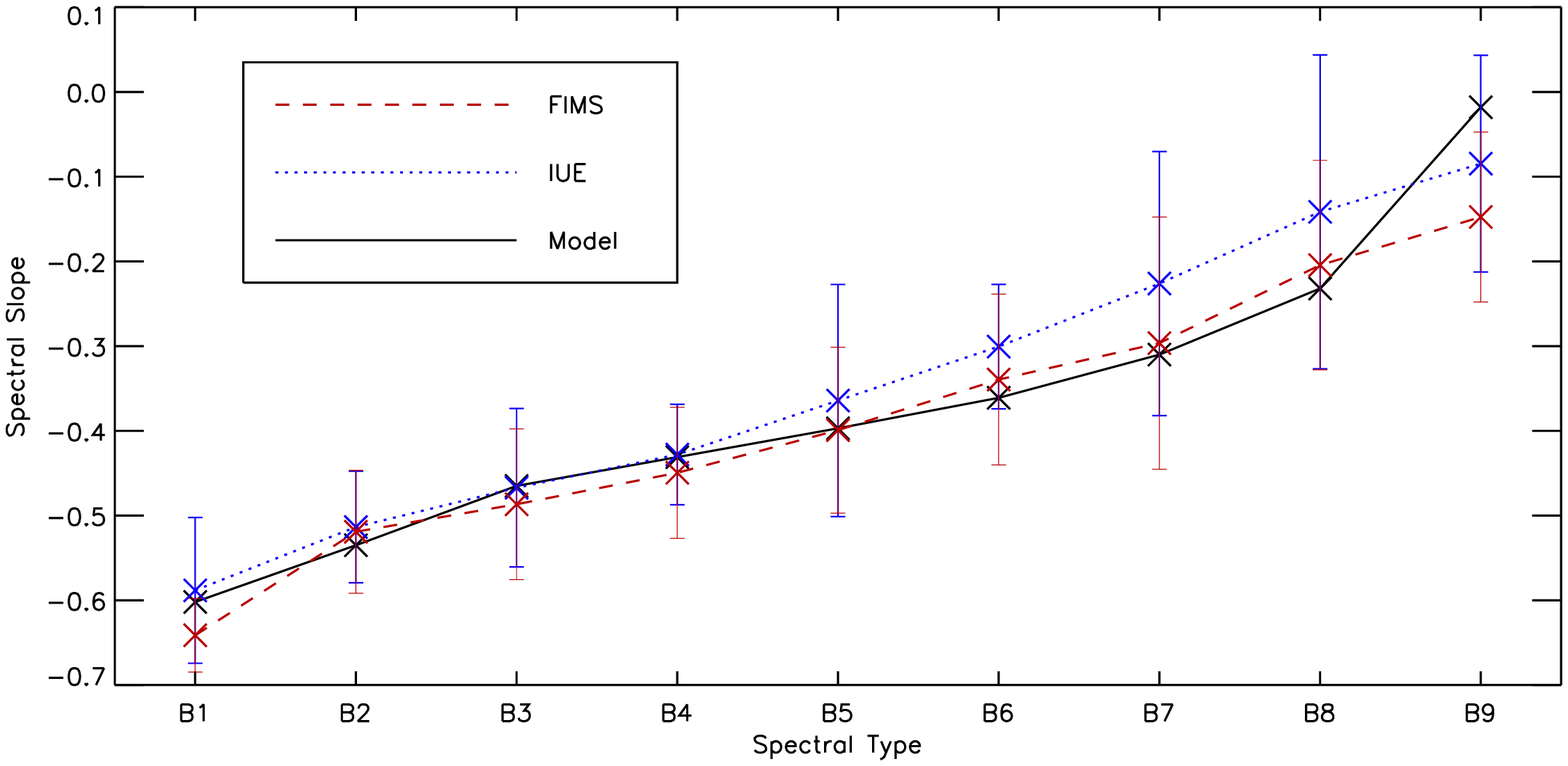}
 \end{center}
 \caption{Variation of the spectral slopes in the FUV passband for
 the stellar spectral type from B1 to B9 .
 The dashed red, dotted blue, and solid black  lines
 indicate the spectral slope variations for the \textit{FIMS}, the \textit{IUE},
 and the Castelli $\&$ Kurucz model \citep{cas03} spectra, respectively. \label{fig:slope_var}}
\end{figure*}

\begin{figure}
 \begin{center}
  \includegraphics[width=8cm]{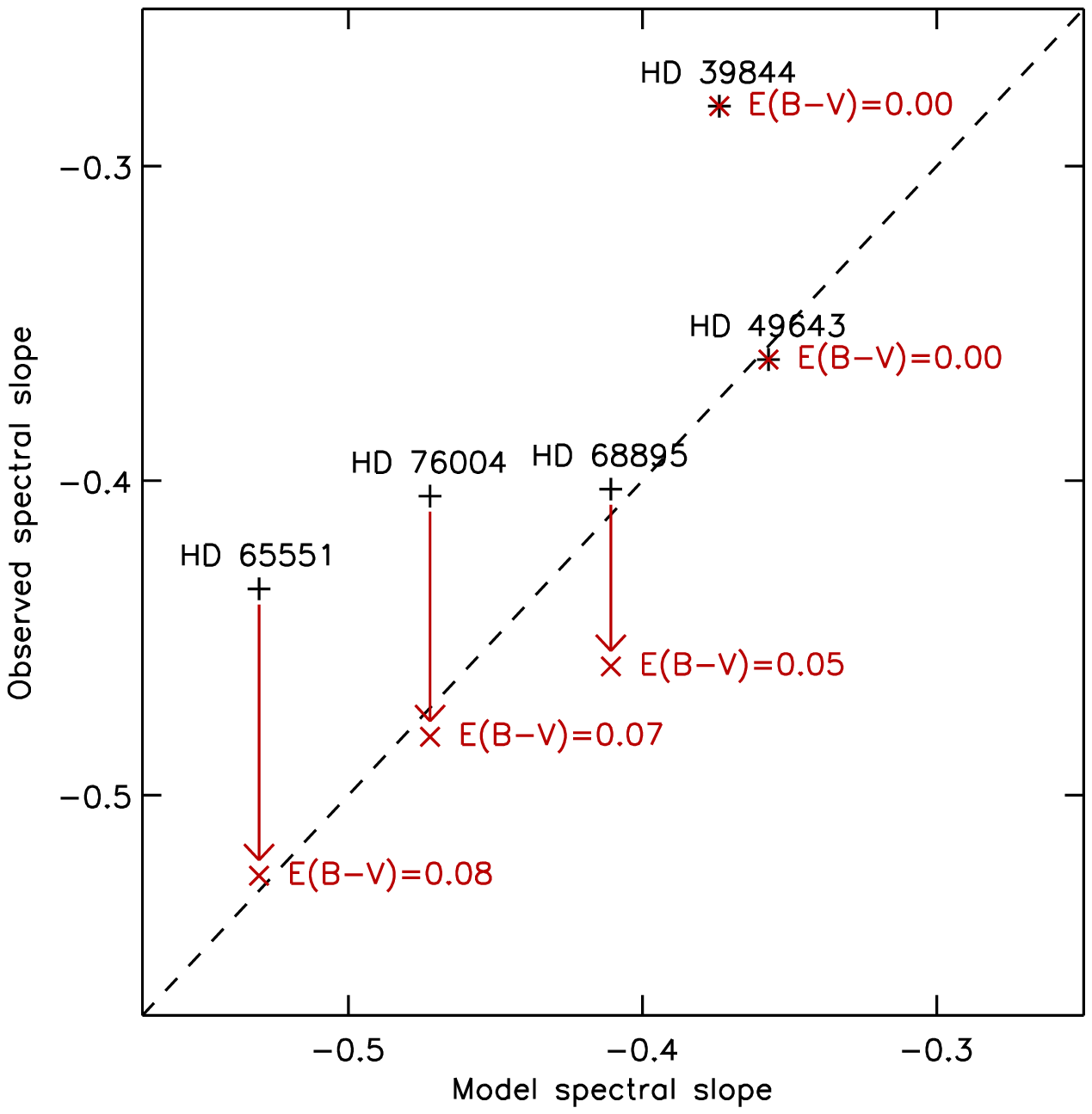}
 \end{center}
 \caption{Comparison of spectral slopes between observations
 and models for several example stars. The black plus (+) symbols indicate
 the spectral slopes before reddening correction, and the red cross ($\times$)
 symbols indicate those after correction. The dashed black line indicates
 the one-to-one correspondence between the observation and model spectral slopes.
 \label{fig:extinction}}
\end{figure}

Finally, we would like to discuss possible applications of the
\textit{FIMS} catalogue. The FUV wavelength band covered by the
\textit{FIMS} includes many ion lines associated with hot and warm
gases. Hence, as can be seen in Figure \ref{fig:sample}, the
prominent ion absorption lines such as \mbox{C\,{\sc iv}}
$\lambda\lambda$1548, 1551 and \mbox{Si\,{\sc iv}}
$\lambda\lambda$1394, 1403 can be used to confirm the spectral
classes for early-type stars. For instance, strong \mbox{Si\,{\sc
iv}} absorption features shown in Figure \ref{fig:sample}(e)
indicate that HD 84567 is a supergiant or giant star with a higher
mass-loss rate rather than a subgiant \citep[see][]{bia02}. On the
other hand, the measured slope (-0.67) after reddening correction is
consistent with -0.66 estimated from the Skiff's spectral type of
B0IV. Another example is HD 65176, which is classified to be A0 in
the SIMBAD database. However, the deep absorption features shown in
Figure \ref{fig:sample}(f) suggest that the star is an early B-type
as assigned to be B1.5Ib/IIep in
\citet{ski14}\footnote{http://vizier.u-strasbg.fr/viz-bin/VizieR?-source=B/MK}.
Furthermore, we note that the spectral shape changes significantly
in this wavelength band as the spectral type changes for the
early-type stars. We, therefore, used the spectral classification of
\citet{ski14} in this paper. As discussed below, this property can
be used to confirm or identify the spectral classes and even
estimate the interstellar extinction in UV wavelengths \citep[e.g.,
the FUV rise;][]{fit88}, which cannot be inferred from optical and
near infrared observations, if the spectral type of the target star
is correctly identified because the interstellar extinction is
strongly wavelength-dependent over this wavelength range.

First, let us define the spectral slope.  The spectral slope is
calculated from a linear fit over the whole wavelength range from
1370 {\AA} to 1710 {\AA}, as shown in Figure \ref{fig:slope_ex} for
the example star, HD 214680. It is defined as follows:
\begin{equation}
   {\rm spectral \; slope} = (F_{1710} - F_{1370})/F_{1540},
\end{equation}
where $F_{1710}$, $F_{1370}$, and $F_{1540}$ are the flux values
obtained from the linear fit at wavelengths of 1710, 1370, and 1540
{\AA}, respectively. We obtained the average spectral slope for each
stellar spectral type based on the \citet{ski14} from B1 to B9 with
$\sim$20 stars for each spectral type. The \textit{E(B-V)} colour
excess was estimated using the intrinsic colour index
\textit{(B-V)$_{0}$} of \citet{fit70} and the observed colour index
\textit{(B-V)} of the SIMBAD database. We selected only those stars
with colour excess \textit{E(B-V)} $<$ 0.05, corresponding to an
optical depth of 0.35 at 1540 {\AA}, and performed reddening
correction using the colour excess. The extinction law was adopted
from \citet{wei01} with R$_V$ = 3.1. We compared the spectral slopes
estimated from the \textit{FIMS} to those of the \textit{IUE} and a
stellar synthetic model. The theoretical model spectra were
calculated through interpolation on a grid of the Kurucz model
\citep{cas03} using the effective temperature and gravity
calibration given in \citet{str81}. Figure \ref{fig:slope_var} shows
the results of the spectral slopes. The three spectral slopes
estimated from the \textit{FIMS}, \textit{IUE}, and a theoretical
model are in good agreement, indicating that the spectral slope is
steeper in earlier spectral types, as expected. The statistical
errors, especially the large ones, may stem from the
misidentification of the spectral type. The sub-classification of
the spectral type contributes in part to the error. A large source
of the uncertainty would be the SIMBAD's photometry and colors.
There are additional sources of uncertainty to the spectral slope.
We assumed a constant $R_V$, which in fact varies from a sightline
to sightline, in performing reddening correction. Another
non-negligible source might be the uncertainty in the calibration of
the effective temperature and gravity. The calibration used in this
paper is based on evolutionary models and thereby subject to their
intrinsic errors \citep[e.g.,][]{her92}.

Figure \ref{fig:extinction} shows some examples of how the spectral
slopes in Figure \ref{fig:slope_var} can be used for the
identification of the spectral type. In Figure \ref{fig:extinction},
the black plus (+) symbols indicate the spectral slopes before
reddening correction and the red cross ($\times$) symbols indicate
those after correction. The sources in the figure were arbitrarily
chosen. We see that the spectral slopes of HD 65551, HD 76004 and HD
49643 are consistent with the model spectral slopes when the
reddening correction is taken into account. On the other hand, the
spectral slopes of HD 68895 and HD 39844 are very different from the
corresponding model spectral slopes. The spectral slope of HD 68895
after correction, estimated from the \textit{FIMS} spectrum, is
-0.46 while the model spectral slope corresponding to the spectral
type of B5V \citep{hou78} is -0.41. This spectral slope of -0.46 is
more consistent with B3V \citep{cuc76} or B4V than B5V if the
reddening correction is correct. If the spectral type of B5V is
correct, the colour excess need to be changed to
\textit{E(B-V)}=0.01, or R$_V$ should be larger than 3.1. In the
same way, the spectral slope of HD 39844 indicates that its spectral
type is more consistent with B7V or B8V than with the spectral type
of B6V \citep{cuc77} if the interstellar extinction is negligible.
On the other hand, if the spectral type of B6V is correct, the
colour excess needs to be changed to \textit{E(B-V)}=0.08, or R$_V$
should be smaller than 3.1 for \textit{E(B-V)} to be less than 0.08.

\section{Concluding Remarks}

We extracted the \textit{L}-band (1370--1710 {\AA}) spectra for 532
stars observed using the \textit{FIMS} during its mission lifetime
of one and a half years, which covered approximately 84\% of the
sky. Of these stars, 323  were also observed by the \textit{IUE}
with higher signal-to-noise ratios than the \textit{FIMS}; hence,
these data were used to validate the \textit{FIMS} spectra. The
remaining 209  stars were compiled as a catalogue: 70 stars were
observed by the \textit{FIMS} for the first time and 139
 stars were observed previously by \textit{UVSST}
and/or \textit{SKYLAB}. We included these 139 stars in the catalogue
because we believe the \textit{FIMS} spectra provide better spectral
resolution and/or higher reliability than the previous observations.
The catalog contains only the stars whose spectra meet the criterion
of average SNRs per angstrom higher than 3.0. The full spectra of
the 209  catalogue stars observed by the \textit{FIMS} can be
downloaded from the KASI website\footnote{http://ysjo.kasi.re.kr}.

\section*{Acknowledgments}

FIMS/SPEAR is a joint project of KAIST (Korea), KASI (Korea), and UC
Berkeley (USA), funded by the Korean MOST and NASA (grant
NAG5-5355).  K.-W. Min acknowledges the support by the National
Research Foundation of Korea through its Grant No. NRF 2012M1A3A4
A01056418.

\bsp

\label{lastpage}

\end{document}